\documentclass[11pt,a4paper]{article}
 
\usepackage{amsmath,amsthm,amssymb}
\usepackage{graphics,graphicx}
\usepackage{epsfig}
\usepackage{multicol}
\usepackage{color}
\makeatletter
\@addtoreset{equation}{section}
\makeatother
\renewcommand{\theequation}{\thesection.\arabic{equation}}

\usepackage{soul}

\usepackage{braket}

\begin{document}

\begin{flushright}
\end{flushright}
\begin{center}
\LARGE{\bf  Which quantum theory must be reconciled with gravity? (And what does it mean for black holes?)}
\end{center}

\begin{center}
\large{\bf Matthew J. Lake} ${}^{a,b}$\footnote{matthewj@nu.ac.th}
\end{center}
\begin{center}
\emph{ ${}^a$ The Institute for Fundamental Study, ``The Tah Poe Academia Institute", \\ Naresuan University, Phitsanulok 65000, Thailand \\}
\emph{ ${}^b$ Thailand Center of Excellence in Physics, Ministry of Education, Bangkok 10400, Thailand \\}
\vspace{0.1cm}
\end{center}

\date{\today}


\begin{abstract}

We consider the nature of quantum properties in non-relativistic quantum mechanics (QM) and relativistic quantum field theories, and examine the connection between formal quantization schemes and intuitive notions of wave-particle duality. Based on the map between classical Poisson brackets and their associated commutators, such schemes give rise to quantum states obeying canonical dispersion relations, obtained by substituting the de Broglie relations into the relevant (classical) energy-momentum relation. In canonical QM, this yields a dispersion relation involving $\hbar$ but not $c$, whereas the canonical relativistic dispersion relation involves both. Extending this logic to the canonical quantization of the gravitational field gives rise to loop quantum gravity, and a map between classical variables containing $G$ and $c$, and associated commutators involving $\hbar$. This naturally defines a ``wave-gravity duality", suggesting that a quantum wave packet describing {\it self-gravitating matter} obeys a dispersion relation involving $G$, $c$ and $\hbar$. We propose an ansatz for this relation, which is valid in the semi-Newtonian regime of both QM and general relativity. In this limit, space and time are absolute, but imposing $v_{\rm max} = c$ allows us to recover the standard expressions for the Compton wavelength $\lambda_C$ and the Schwarzschild radius $r_S$ within the same ontological framework. The new dispersion relation is based on ``extended" de Broglie relations, which remain valid for slow-moving bodies of {\it any} mass $m$. These reduce to canonical form for $m \ll m_P$, yielding $\lambda_C$ from the standard uncertainty principle, whereas, for $m \gg m_P$, we obtain $r_S$ as the natural radius of a self-gravitating quantum object. Thus, the extended de Broglie theory naturally gives rise to a unified description of black holes and fundamental particles in the semi-Newtonian regime.

\end{abstract}

\tableofcontents


\section{Introduction} \label{Sec.1}

The Compton wavelength $\lambda_C$ gives the minimum radius within which the mass of a particle may be localized due to relativistic quantum effects, neglecting the particle's self-gravity, while the Schwarzschild radius $r_S$ gives the maximum radius within which the mass of a black hole may be localized due to classical relativistic gravity, neglecting quantum effects. In a mass-radius diagram, the two lines intersect near the Planck point $(m_P,l_P)$, where $m_P$ denotes the Planck mass and $l_P$ the Planck length, so that relativistic-quantum-gravitational effects become significant in this region. In addition, the equivalence of the Compton and Schwarzschild radii close to the Planck scales suggests that a final theory of quantum gravity (QG) should yield a unified description of fundamental particles and black holes. Nonetheless, how such a unification could be achieved may be considered an outstanding problem for both elementary particle physics, and the physics of black holes, even in the absence of a final theory. 

Since canonical {\it non-gravitational} QM and quantum field theories (QFTs) are based on the concept of wave-particle duality, encapsulated in the de Broglie relations, these may break down close to $(m_P,l_P)$. However, it is unclear what (if any) physical interpretation can be given to quantum particles with energies $E \gg m_Pc^2$, since these correspond to ``matter waves" with sub-Planckian wavelengths $\lambda \ll l_P$ and time periods $\tau \ll t_P$ ($t_P = l_P/c$), according to canonical quantum dispersion relations. On the other hand, ``particles" with rest masses $m > m_P$ have sub-Planckian Compton wavelengths, $\lambda_C < l_P$, but super-Planckian Schwarzschild radii, $r_S > l_P$, and may be interpreted as black holes. We therefore propose corrections to the standard de Broglie relations, which are valid in the semi-Newtonian limit, allowing them to be extended into the region $m \gg m_P$. 

Here, the semi-Newtonian theory is defined as the Newtonian limit of general relativity, in which space and time are absolute but there exists a cosmic speed limit $v_{\rm max}=c$, which must be imposed ``by hand" on the quantum sector. Though, from the viewpoint of a final theory, this is {\it clearly} inadequate, it has the advantage of allowing us to describe key phenomenological features of spherically symmetric systems on both halves on the mass-radius diagram, $m \leq m_P$ and $m \geq m_P$, within a unified ontological framework. In other words, using the same set of assumptions about the nature of space and time, the finite speed of light, and the form of the quantum dispersion relations, we recover well-known features of both canonical QM and classical gravity. These assumptions gives rise to both a modified Schr{\" o}dinger equation and a modified expression for the Compton wavelength. Both reduce to the standard forms for $m \ll m_P$, as required for consistency, but, crucially, also allow us to recover the standard expression for the Schwarzschild radius for $m \gg m_P$, using standard ``quantum" arguments. Thus, we interpret the additional terms in the modified de Broglie relations, which involve all three fundamental constants, $G$, $c$ and $\hbar$, as representing the self-gravitation of the wave packet in the semi-Newtonian background. 

The structure of this paper is as follows. In Sec. \ref{Sec.2}, we review the basic arguments for the ``quantum" nature of non-relativistic point-particles and relativistic fields, focussing on the equivalence of wave-particle duality, as expressed via the canonical dispersion relations, and formal quantization schemes. In Sec. \ref{Sec.3.1}, we briefly review Newtonian gravity and general relativity, with special emphasis on their application to spherically symmetric bodies (i.e., particles and black holes). The gravitational implications of the the semi-Newtonian approximation are discussed in \ref{Sec.3.2}. Sec. \ref{Sec.3.3} considers two previous attempts to incorporate quantum effects into gravitational theories, the Schr{\" o}dinger-Newton equation and Loop Quantum Gravity (LQG). Their relations to the extended de Broglie theory, and to the problem of obtaining a unified description of fundamental particles and black holes, are also discussed. In Sec. \ref{Sec.3.4}, the quantum mechanical implications of the semi-Newtonian approximation are considered and the conceptual framework for the extended de Broglie theory is defined. The ansatz for the extended relations is motivated by considering the asymptotic black hole regime in Sec. \ref{Sec.4.1} and the full relations are presented in Sec. \ref{Sec.4.2}. In Sec. \ref{Sec.5}, we determine the equations of motion (EOM) for the quantum state and the associated Hamiltonian and momentum operators, in both the particle and black hole regimes. The implications of the extended theory for the Hawking temperature are discussed in Sec. \ref{Sec.6}, and alternative unification schemes for black holes and fundamental particles, based on Generalized Uncertainty Principle (GUP) phenomenology, are considered in Sec. \ref{Sec.7}, with special emphasis on how these relate to the ideas discussed in Secs. \ref{Sec.2}-\ref{Sec.6}. A brief summary our our main results is given in Sec. \ref{Sec.8}. For reference, the four postulates of canonical QM are listed in the Appendix. 

Thus, the results presented in Secs. \ref{Sec.2}-\ref{Sec.3} are mostly pedagogical, while those presented in Secs. \ref{Sec.4}-\ref{Sec.6} are based primarily on results obtained in \cite{Lake:2015pma}, in which the extended de Broglie theory was first introduced. However, the purpose of the present work is not simply to review the results of previous studies, but to propose a provisional answer to the questions posed in the title, ``Which quantum theory must be reconciled with gravity? (And what does it mean for black holes?)", which may be considered important open questions in black hole physics. For this reason, the material presented in Secs. \ref{Sec.2}-\ref{Sec.3} is chosen to highlight the steps required to obtain a unified ontology for the description of black holes and fundamental particles. Since this {\it necessarily} entails important modifications to our existing notions of what makes a theory ``quantum", it leads naturally to the extended de Broglie theory as a {\it provisional} scheme for unification in the semi-Newtonian limit. Throughout, we explore the relation between this theory and existing models of modified quantum theory $-$ relativistic, non-relativistic and semi-relativistic $-$ motivated by gravitational considerations. In each case, we attempt to clarify the theoretical assumptions underpinning different approaches and to identify which (if any) of the physical assumptions on which different models are based may be in contradiction with one another. In particular, special emphasis is given to particle-black hole unification schemes arising in the context of GUP phenomenology, which are considered in detail in Sec. \ref{Sec.7}.

\section{What makes a theory ``quantum"?} \label{Sec.2}

The essence of quantum theory is wave-particle duality. Though, philosophically, a somewhat slippery concept \cite{Bell:1987hh}, it is encapsulated mathematically in a very precise form via the de Broglie relations \cite{Rae00}
\begin{eqnarray} 
\label{deBroglie}
E = \hbar\omega \, , \quad \quad  \vec{p} = \hbar\vec{k} \, . 
\end{eqnarray} 
In 1924, de Broglie's great insight was to relate the constants of motion of a classical point particle, energy $E$ and momentum $\vec{p}$, with properties hitherto associated only with waves; frequency $f$ and wavelength $\lambda$, or, equivalently, angular frequency $\omega = 2\pi f$ and wave number $k = 2\pi/\lambda$. This was made possible by Planck's earlier discovery of new fundamental constant with dimensions of action, $\hbar = h/(2\pi) = 1.055 \times 10^{-34} {\rm Js}$, which proved necessary to explain the observed spectrum of black body radiation. The quantization of absorbed and emitted radiation, according to Eq. (\ref{deBroglie}), was able to account for the suppression of high-energy modes that otherwise result in an ``ultraviolet catastrophe" \cite{UV}. 

The breakthrough leading, ultimately, to the development of modern quantum mechanics (QM) and quantum field theory (QFT) came when de Broglie proposed that the relations (\ref{deBroglie}) were applicable to {\it all} forms of matter and radiation. Thus, the foundation of {\it any} quantum theory is the quantum dispersion relation, which is obtained by substituting (\ref{deBroglie}) into the classical relation between $E$ and $\vec{p}$. In general, this may be relativistic or non-relativistic and may apply to objects with zero or nonzero rest mass. Typically, the latter are considered ``particles" in the classical theory and acquire wave-like characteristics via quantization, whereas the former are considered classically as ``waves", which acquire particle-like properties via (\ref{deBroglie}). The equation of motion (EOM) for the quantum state must satisfy the dispersion relation obtained by combining the classical energy-momentum relation with the de Broglie relations and be invariant under the symmetries of the corresponding classical theory.

For massive particles, the quantum dispersion relation involves $\hbar$ and will also include $c$, the speed of light, if the theory is relativistic. Roughly speaking, we may say that a theory describing massive particles is quantum (meaning it incorporates wave-particle duality) and non-relativistic if its dispersion relation contains, $\hbar$ but not $c$, and if the resulting EOM are invariant under rotations, translations in time and space, and local Gallilean boosts. Likewise, a theory of massive particles is quantum and relativistic if its dispersion relation contains both $\hbar$ and $c$, and if the resulting EOM are invariant under the action of the Poinc{\' a}re group, comprising rotations, translations and local Lorentz boosts.

\subsection{Wave-particle duality in Newtonian space -- without Newtonian gravity} \label{Sec.2.1}
The non-relativistic energy-momentum relation is 
\begin{eqnarray} 
\label{non_rel_E-p}
E = \frac{p^2}{2m} + V \, ,
\end{eqnarray} 
where $V(\vec{x},t)$ represents an external classical potential, not generated by the particle itself. The quantum dispersion relation is, therefore,  
\begin{eqnarray} 
\label{non_rel_disp}
\omega = \frac{\hbar k^2}{2m} + V \, .
\end{eqnarray} 
By assuming that a single ``matter wave" mode, corresponding to a single value of the classical momentum $\vec{p}$, is given by $\exp[i(\vec{k}.\vec{x} - \omega t)]$, and that a general quantum state $\psi(\vec{x},t)$ is given by a superposition of modes, the next fundamental breakthrough was made by Schr{\" o}dinger, who obtained the EOM for a non-relativistic quantum system:
\begin{eqnarray} 
\label{Schrodinger}
\hat{H}\psi = i\hbar\frac{\partial \psi}{\partial t}, \quad \quad \hat{H} = \left(-\frac{\hbar^2}{2m}\vec{\nabla}^2 + V\right) \, .
\end{eqnarray} 
For a given $V$, this equation is {\it unique} and, hence, is the only EOM consistent with Eqs. (\ref{deBroglie})-(\ref{non_rel_E-p}). The operator $\hat{H}$ is called the Hamiltonian and its eigenvalues represent the possible results of measurements of the system's energy.  

In the position space representation, the results of position measurements are given by the classical position vector, so that the operators corresponding to classical canonical coordinates are given by 
\begin{eqnarray} 
\label{canonical_coord}
\hat{\vec{x}} = \vec{x}, \quad \quad \hat{\vec{p}} = -i\hbar \vec{\nabla} \, ,
\end{eqnarray} 
with eigenfuctions
\begin{eqnarray} 
\label{posn_space_eigenfn}
\phi(\vec{x},\vec{x}_0) = \delta(\vec{x}-\vec{x}_0), \quad \quad \phi(\vec{x},t) = \exp[i(\vec{k}.\vec{x} - \omega t)] \, .
\end{eqnarray} 
The equivalents of Eqs. (\ref{canonical_coord})-(\ref{posn_space_eigenfn}) in the momentum space representation are obtained by applying the transformation $\vec{x} \leftrightarrow \vec{p}$, so that the position and momentum operators obey the canonical commutation relations,
\begin{eqnarray} 
\label{comm_rel-1}
[\hat{x}^i,\hat{p}_j] = i\hbar \delta^{i}{}_{j} \, ,
\end{eqnarray} 
in either representation. In general, consistency requires that these hold in {\it any} representation, so that (\ref{comm_rel-1}) may be taken as a definition: any pair of operators satisfying the canonical commutation relations are valid representations of $\hat{x}^i$ and $\hat{p}_j$. It is a fundamental postulate of canonical QM that operators representing other dynamical variables bear the same functional relation to these as do the corresponding classical quantities to the classical position and momentum \cite{Rae00}. 

In the Schr{\" o}dinger picture, operators representing physical observables are time-independent, whereas the quantum state $\psi$ is time-dependent, according to 
\begin{eqnarray} 
\label{time_evol_Schrod}
\hat{O} = \hat{O}(\vec{x}), \quad \quad \psi(\vec{x},t) = \psi(\vec{x},0)e^{-i(\hat{H}/\hbar) t} \, .
\end{eqnarray} 
Equivalently, in the Heisenberg picture, operators are time-dependent and states time-independent, so that
\begin{eqnarray} 
\label{time_evol_Heis}
\hat{O}(\vec{x},t) = e^{i(\hat{H}/\hbar)t}\hat{O}(\vec{x},t)e^{-i(\hat{H}/\hbar)t}, \quad \quad \psi = \psi(\vec{x})  \, ,
\end{eqnarray} 
and the quantum EOM becomes
\begin{eqnarray} 
\label{Heisenberg}
\frac{d \hat{O}}{dt} = -\frac{i}{\hbar}[\hat{O},\hat{H}] + \frac{\partial \hat{O}}{\partial t} \, .
\end{eqnarray} 
Equation (\ref{Heisenberg}) takes exactly the same form as the Hamilton-Jacobi equation in classical non-relativistic mechanics \cite{Landau} under the transformation 
\begin{eqnarray} 
\label{Poisson<->commutator-1}
-\frac{i}{\hbar}[\hat{H},\hat{O}] \leftrightarrow \left\{H,f\right\} \, ,
\end{eqnarray} 
where 
\begin{eqnarray} 
\label{Poisson_bracket}
\left\{f,g\right\} = \frac{\partial f}{\partial q^i}\frac{\partial g}{\partial p_i} - \frac{\partial f}{\partial p_i}\frac{\partial g}{\partial q^i} \, ,
\end{eqnarray}
is the classical Poisson bracket for the functions $f(q^i,p_i)$, $g(q^i,p_i)$. (The Schr{\" o}dinger equation may also be related to Hamilton's equation via the transformation $\psi = \psi_0e^{iS/\hbar}$, where $S(q^i,p_i,t)$ is the action.) In classical mechanics, the Poisson bracket for the canonical coordinates is 
\begin{eqnarray} 
\label{Poisson_canon_coord}
\left\{x^i,p_j\right\} = \delta^{i}{}_{j}\, ,
\end{eqnarray}
so that applying the transformation 
\begin{eqnarray} 
\label{Poisson<->commutator-2}
-\frac{i}{\hbar}[\hat{x}^i,\hat{p}_j] \leftrightarrow \left\{x^i,p_j\right\} \, 
\end{eqnarray}
yields (\ref{comm_rel-1}). These results suggest a general {\it quantization scheme}, defined by the following association between classical quantities and the Hermitian operators representing the corresponding QM observables \cite{Dirac}:
\begin{eqnarray} 
\label{quantization_scheme}
\lim_{\hbar \rightarrow 0}\frac{[\hat{O}_1,\hat{O}_2]}{i\hbar} = \left\{O_1,O_2\right\} \, .
\end{eqnarray}

Such an abstract scheme seems far removed from our initial considerations regarding the dual wave/particle nature of quantum states, but follows logically from the definition of wave-particle duality encapsulated in the de Broglie relations (\ref{deBroglie}). The principle of superposition, applied to de Broglie waves, also implies the equivalence of the wave function $\psi$, in real space, and the wave vector $\ket{\psi}$, in a vector space that forms the state space of the theory. Thus, a general quantum state $\ket{\psi}$ may be expressed as a superposition of eigenfunctions of an arbitrary operator $\hat{O}$, which form an arbitrary set of basis vectors in the corresponding Hilbert space \cite{Rae00,Dirac,Ish95}. This, in turn, is key to the interpretation of QM as a probabilistic theory, and the entire formalism of canonical QM can be defined in terms of four basic Postulates that relate its mathematical structure to the outcomes of physical measurements \cite{Ish95}. (For reference, these are listed in the Appendix.) An immediate consequence is the existence of the General Uncertainty Principle, for a pair of arbitrary operators,
\begin{eqnarray} \label{SUP}
\Delta_{\psi}O_1 \Delta_{\psi}O_2 
\geq \frac{1}{2}\sqrt{|\langle \psi|[\hat{O}_1,\hat{O}_1]|\psi\rangle|^2 + |\langle \psi|[\hat{A},\hat{B}]_{+}|\psi\rangle|^2}
\geq \frac{1}{2}|\langle \psi|[\hat{O}_1,\hat{O}_1]|\psi\rangle| \, , 
\end{eqnarray}
where $[\hat{O}_1,\hat{O}_2]$ is the commutator of $\hat{O}_1$ and $\hat{O}_2$, $[\hat{A},\hat{B}]_{+}$ is the anti-commutator of $\hat{A} = \hat{O}_1 - \langle \hat{O}_1\rangle_{\psi}\hat{\mathbb{I}}$ and $\hat{B} = \hat{O}_2 - \langle \hat{O}_2\rangle_{\psi}\hat{\mathbb{I}}$, and 
\begin{eqnarray} 
\label{uncertainty}
\Delta_{\psi}O = \sqrt{\langle \hat{O}^2 \rangle_{\psi} - \langle \hat{O} \rangle_{\psi}^2} \, , \quad \langle \hat{O} \rangle_{\psi} = \bra{\psi}\hat{O}\ket{\psi} \, .
\end{eqnarray}
$\Delta_{\psi}O$ is called the uncertainty and  $\langle \hat{O} \rangle_{\psi}$ the expectation value of $\hat{O}$. Equation (\ref{SUP}) follows directly from the Hilbert space structure of canonical QM via the Schwarz inequality and, setting $\hat{O}_1 = \hat{x}^i$ and $\hat{O}_2 = \hat{p}_j$, we obtain the famous Heisenberg Uncertainty Principle (HUP):
\begin{eqnarray} \label{HUP}
\Delta_{\psi}x^i \Delta_{\psi}p_j \geq \frac{\hbar}{2}\delta^{i}{}_{j}  + \mathcal{O}_{\psi}(\hbar^2) \, .
\end{eqnarray}

Furthermore, it may be shown that operators representing classically conserved quantities may be identified (up to factors of $\hbar$) with the generators of the associated symmetry group. In classical physics, the state space is a manifold and the invariance of an arbitrary state under a given isometry leads to existence of a conserved charge $Q$ via Noether's theorem \cite{Landau2}. In quantum mechanics, the state space is a vector space and the invariance of an arbitrary state under the same isometry gives rise to the canonical commutation relation involving $\hat{Q}$ \cite{Ish95}. 

The discussion above gives a brief account of the logical development of the formalism of canonical non-relativistic QM. For our current purposes, the key point is that, no matter how abstract this formalism appears, its fundamental physical {\it root} is the association of particle and wave properties according to (\ref{deBroglie}). The precise form of the dispersion relation (\ref{non_rel_disp}), EOM (\ref{Schrodinger}), commutation relations (\ref{SUP}), and state space (Hilbert space) structure, follow from the way in which the de Broglie relations are combined with the energy-momentum relation for a point-particle in Newtonian mechanics (\ref{non_rel_E-p}). Crucially, this approach {\it ignores} the effect of the particle's self-gravity, even in the Newtonian limit. 


\subsection{Wave-particle duality in Minkowski space -- without general relativity} \label{Sec.2.1}

In the relativistic case, the classical energy-momentum relation for a {\it free} particle is 
\begin{eqnarray} 
\label{rel_E-p}
E^2 = p^2c^2 + m^2c^4, \quad \quad E = \pm \sqrt{p^2c^2 + m^2c^4} \, .
\end{eqnarray}
Combining this with (\ref{deBroglie}), the quantum dispersion relation is 
\begin{eqnarray} 
\label{rel_disp}
\omega^2 = c^2(k^2 + k_C^2), \quad \quad \omega = \pm c\sqrt{k^2 + k_C^2} \, ,
\end{eqnarray}
where 
\begin{eqnarray} 
\label{Compton}
k_C = \frac{2\pi}{\lambda_C}, \quad \quad  \lambda_C = \frac{h}{mc} \, ,
\end{eqnarray}
and $\lambda_C$ is the Compton wavelength. This is the length-scale that can be naturally associated with the particle's rest mass using the constants $h$ and $c$ and may be interpreted as the minimum possible radius of a quantum ``particle" of mass $m$. (The quantity $k_C^{-1} = \hbar/(mc)$ is known as the reduced Compton wavelength.) In this case, the corresonding EOM is {\it not} unique, and depends on several factors, including:
\begin{itemize}

\item The explicit form in which the quantum dispersion relation is written, i.e. as on the left-hand or right-hand side of Eq. (\ref{rel_disp}).

\item The gauge symmetries -- in addition to the Poinc{\' a}re symmetry of the Minkowski space background -- under which the action of the corresponding classical theory is invariant. 

\end{itemize}
A fundamental difference between this and the non-relativistic case is the existence of two solution branches, which seem to give rise to particles with both positive and negative energy. Furthermore, although the two forms of the energy-momentum relation given in (\ref{rel_E-p}) are classically equivalent, this equivalence does not extend to the corresponding quantum EOM, obtained via the substitutions $E \rightarrow \hat{H}$, $p \rightarrow \hat{p}$. In other words, the two forms of the dispersion relation given on the left-hand and right-hand sides of Eq. (\ref{rel_disp}) are quantum mechanically {\it inequivalent}.

An example of an EOM satisfying the left-hand (``squared") form of the dispersion relation is the Klein-Gordon equation \cite{Peskin:1995ev},
\begin{eqnarray} 
\label{KG}
\frac{1}{c^2}\frac{\partial^2 \psi}{\partial t^2} - \vec{\nabla}^2\psi + \frac{m^2c^2}{\hbar^2}\psi = 0 \, ,
\end{eqnarray}
which may be generalized to describe a wave function moving in a potential $V(\psi)$ by adding the term $\partial V/\partial \psi$. Equation (\ref{KG}) obeys both Poinc{\' a}re symmetry and the local $U(1)$ gauge symmetry of the complex scalar $\psi$, which gives rise to a conserved electric current. It describes spinless, electrically charged, relativistic quantum particles, but is {\it not} the EOM for a fully self-consistent theory of a quantum one-particle state. Though the mathematical subtleties are complex, the physical reason for this is simple: no such theory exists. In fully self-consistent theories (QFTs), negative energy states given by the right-hand side of (\ref{rel_E-p}) may be reinterpreted as positive energy particles under charge inversion ($+ \leftrightarrow -$), so that relativistic quantum theories predict the existence of anti-matter. States with $E \gtrsim mc^2$ then lead to the the production of particle--anti-particle pairs, giving rise to multi-particle states that conserve electric charge \cite{Peskin:1995ev}. 

It is straightforward to show that the classical inequality $E \gtrsim mc^2$ corresponds to $\lambda \lesssim \lambda_C$ in the quantum picture, so that pair-production occurs when the size of a local field oscillation becomes smaller than the Compton wavelength. This gives rise to an effective cut-off for the spread of the wave packet in the non-relativistic theory, $\Delta_{\psi}x \gtrsim \lambda_C$, so that the HUP (\ref{HUP}) implies 
\begin{eqnarray} 
\label{Compton-HUP}
(\Delta_{\psi}x)_{\rm min} \approx \lambda_C, \quad \quad (\Delta_{\psi}p)_{\rm max} \approx mc \, .
\end{eqnarray}
Hence, while the de Broglie wavelength of an object marks the length-scale at which non-relativistic quantum effects become important for its description, and the classical concept of a particle gives way to the wave packet, the Compton wavelength marks the point at which relativistic quantum effects become significant and the concept of a single wave packet corresponding to a state in which particle number remains fixed breaks down. This acts as an effective minimum width because, for de Broglie modes with $\lambda \lesssim \lambda_C$, we must switch to a field description in which particle creation occurs in place of further spatial localization.

By contrast with Eq. (\ref{KG}), the right-hand (``square root") form of the quantum dispersion relation gives rise, {\it uniquely}, to the Dirac equation, from which the original prediction of anti-matter was derived \cite{Rae00,Dirac,Peskin:1995ev}. This may be written in a manifestly Lorentz invariant form as
\begin{eqnarray} 
\label{Dirac}
\left(\gamma^{\mu}\frac{\partial }{\partial  x^{\mu}} - \frac{mc}{\hbar}\right)\psi = 0 \, ,
\end{eqnarray}
where $\mu \in \left\{0,1,2,3\right\}$ are space-time indices and $\gamma^{\mu}$ denotes a set of $4 \times 4$ matrices 
\[\gamma^{0} = \left[ \begin{array}{cc} \label{Dirac_mats}
0                        &           \mathbb{I}_2                      \\
\mathbb{I}_2      &            0          
\end{array} \right], \ \ \ 
\gamma^{i} = \left[ \begin{array}{cc} 
0                        &           \sigma^{i}                      \\
 \sigma^{i}         &            0      
\end{array} \right] \] 
where $\mathbb{I}_2$ is the $2 \times 2$ identity matrix and $\sigma^{i}$ are the Pauli matrices \cite{Rae00,Peskin:1995ev}. The Dirac equation describes spin-$1/2$, electrically charged, relativistic quantum particles. Crucially, it also enables us to explain the (otherwise mysterious) property of quantum mechanical spin as a necessary consequence of the union of quantum mechanics and special relativity.

It is straightforward to show that ``squaring" the Dirac equation (\ref{Dirac}) gives the Klein-Gordon equation (\ref{KG}) and that Taylor expanding this -- using an appropriate definition of the ``square root" of a differential operator -- yields the Schr{\" o}dinger equation (\ref{Schrodinger}). In fully consistent QFTs, the Klein-Gordon equation also reemerges as the EOM governing the components of {\it all} free (i.e. non-interacting) quantum fields. The solutions of the free field EOM are used to build the state space -- in this case, a Fock space \cite{Peskin:1995ev} -- of the theory, since these form a basis that spans the space. Hence, while integer spin particles may obey {\it any} EOM that is consistent with the ``squared" form of the quantum dispersion relation, together with the other requirements of the theory, such as gauge invariance, etc., particles with half-integer spin always obey the Dirac equation, or its generalization to include interactions with electromagnetic field \cite{Rae00,Peskin:1995ev}. However, since only {\it two} quantum mechanically inequivalent forms of the dispersion relation exist, there are only {\it two} types of quantum relativistic particle: bosons and fermions.

In the field theory approach, the classical field variables become quantum operators, so that, for example, a scalar field $\phi(x)$ and its canonical momentum $\pi(x)$ satisfy the relation \cite{Peskin:1995ev}
\begin{eqnarray} 
\label{canon_comm_rel_QFT}
[\hat{\phi}(x),\hat{\pi}(x)] = i\hbar\delta^{3}(x-x')\, .
\end{eqnarray}
This is equivalent to the map
\begin{eqnarray} 
\label{canon_quant_QFT}
\lim_{\hbar \rightarrow 0}\frac{[\hat{\phi}(x),\hat{\pi}(x)]}{i\hbar} = \left\{\phi(x),\pi(x)\right\} \, ,
\end{eqnarray}
where 
\begin{eqnarray} 
\label{Poisson_CFT}
\left\{\phi(x),\pi(x)\right\} = \delta^{3}(x-x')
\end{eqnarray}
is the classical Poisson bracket. Thus, equations like (\ref{canon_comm_rel_QFT}) represent the field theory analogue of the canonical quantization of point particles, (\ref{quantization_scheme}).

The discussion in this subsection gives a brief account of the logical development of relativistic QM and a very brief introduction to some of the basic concepts of QFT. We have seen that anti-matter, bosons and fermions, the existence of the Compton wavelength and pair-production arise as necessary consequences of any theory that successfully incorporates both wave-particle duality -- encapsulated in the de Broglie relations (\ref{deBroglie}) -- and local Lorentz invariance. Though the field replaces the particle as the fundamental object to be ``quantized", and multiple excitations are interpreted as multi-particle states \cite{Peskin:1995ev}, the time-evolution of each excitation remains consistent with the point-particle dispersion relation (\ref{rel_disp}). 

For our current purposes the key point is that, no matter how complex or abstract the mathematical formulation of realistic QFTs -- including the $SU(3) \times SU(2) \times U(1)$ gauge theories of the Standard Model of particle physics -- become, their basic architecture {\it must} be compatible with the dispersion relation (\ref{rel_disp}), which follows directly from combining the de Broglie relations with the energy-momentum relation for a point-particle in special relativity. Again, this approach {\it ignores} the effect of the particle's self-gravity, which, in the relativistic regime, must be described by Einstein's theory of general relativity.  


\section{What makes a theory gravitational?} \label{Sec.3}

By our previous logic, a crude definition of a gravitational theory is any set of EOM involving Newton's constant $G$ that reduce to Einstein's field equations (EFEs) in an appropriate limit. These, in turn, reduce to the Newtonian theory of gravity in the weak field approximation \cite{Dirac2}. The EFEs include both $G$ and $c$, and incorporate both relativistic and gravitational effects. More specifically, we may say that they describe the regime in which relativistic-gravitational effects become important.

\subsection{Newtonian and Einstein gravity -- without wave-particle duality} \label{Sec.3.1}

We now briefly review Newtonian gravity and the the general theory of relativity, with special emphasis on their application to spherically symmetric systems. Specifically, we show how the Newtonian gravitational potential arises as the weak field limit of the unique, static, spherically symmetric vacuum solution of the EFEs, the Schwarzschild solution \cite{Dirac2}. 

\subsubsection{Newtonian gravity for spherically symmetric objects} \label{Sec.3.1.1}

In spherical polar coordinates, Newton's law of gravity gives the gravitational potential energy (per unit mass) of a body of mass $M$, or, equivalently, the force between two spherically symmetric bodies with masses $M$ and $M'$:
\begin{eqnarray} 
\label{Newton}
\Phi(r) = -\frac{GM}{r}, \quad \quad  \vec{F} = -\frac{GM M'}{r^2}\hat{\underline{r}} \, . 
\end{eqnarray}
This is consistent with Newton's second law of motion, $\vec{F} = M\vec{a}$, since $\vec{a} = -\vec{\nabla}\Phi$. The potential is the Green's function \cite{Boas} solution of Poisson's equation,
\begin{eqnarray} 
\label{Poisson_eq}
\vec{\nabla}^2\Phi =  4\pi G\rho \, ,
\end{eqnarray}
where $\rho$ is the mass density of a gravitating body. This is invariant under rotations, translations and Gallilean boosts, as required by the background of absolute space and time. Though it is straightforward to set $V(r) = \Phi(r)$ in Schr{\" o}dinger's equation, this models a quantum wave packet moving in an external (classical) gravitational field, and cannot account for the self-gravity of the quantum ``particle". At present it is unclear whether such an effect can be successfully incorporated into canonical QM, though one semi-classical attempt -- the Schr{\" o}dinger-Newton equation -- is described in Sec. \ref{Sec.3.3}.

\subsubsection{General relativity and the Schwarzschild solution} \label{Sec.3.1.2}

The general theory of relativity is based on the action \cite{Dirac2}
\begin{eqnarray} 
\label{Einstein-Hilbert}
S =  \frac{c^4}{16\pi G}\int R\sqrt{-g}d^4x + \int \mathcal{L}_M \sqrt{-g}d^4x\, .
\end{eqnarray}
The first integral is the Einstein-Hilbert term, which describes the vacuum theory, and $\mathcal{L}_M$ is the matter Lagrangian. 
Variation with respect to the metric $g_{\mu\nu}$ gives the EFEs
\begin{eqnarray} 
\label{Einstein}
G_{\mu\nu} \equiv R_{\mu\nu} - \frac{1}{2}Rg_{\mu\nu} = \frac{8\pi G}{c^4}T_{\mu\nu} \, ,
\end{eqnarray}
where the energy-momentum tensor for the matter sector is defined implicitly via
\begin{eqnarray} 
\label{T^{munu}}
\delta S_M = -\frac{1}{2} \int d^4x \sqrt{-g} T^{\mu\nu} \delta g_{\mu\nu} \, .
\end{eqnarray}
$G_{\mu\nu}$ is called the Einstein tensor, $R_{\mu\nu} = R^{\alpha}{}_{\mu\alpha\nu}$ is the Ricci tensor, $R = R^{\mu}{}_{\mu}$ is the scalar curvature, and $R_{\mu\nu\alpha\beta}$ is the Riemann curvature tensor \cite{Dirac2}. This determines the deviation of geodesics due to the curvature of space-time, induced by the presence of matter \cite{Misner:1974qy}. 

In the static, weak field limit, the vacuum EFEs reduce to \cite{Dirac2,Misner:1974qy}
\begin{eqnarray} 
\label{weak_field-1}
\frac{1}{\sqrt{-g}}\frac{\partial}{\partial x^i}\left(\sqrt{-g}g^{ij}\frac{\partial g_{00}}{\partial x^i}\right) = 0 \, ,
\end{eqnarray}
where $i,j$ denote spatial coordinates, which is simply the covariant version of Laplace's equation (Poisson's equation with $\rho = 0$). As $g_{00}/c^2$ must be close to unity, we may set 
\begin{eqnarray} 
\label{weak_field-2}
g_{00}/c^2 = 1 + 2\Phi/c^2 \, , 
\end{eqnarray}
where $\Phi/c^2 \ll 1$, giving $g_{00}^{1/2}/c = 1+\Phi/c^2$. The function $\Phi/c^2$ also obeys Laplace's equation, but, as it does {\it not} correspond to the flat space solution, must be identified with the Green's function solution of Poisson's equation and hence with the Newtonian potential $-GM/r$ (\ref{Newton}). In this way, $c$ explicitly ``drops out" of the EFEs in the Newtonian approximation, as expected. Nonetheless, the derivation of the Newtonian potential from the weak field limit of Einstein's theory shows that it is {\it compatible} with a cosmic speed limit $c$, which may be re-introduced ``by hand", giving rise to a more accurate semi-Newtonian description. 

Since all quantities in Eq. (\ref{Einstein}) are expressed in terms of tensors, the EFEs are invariant under general coordinate transformations (GCTs), a property also known as general covariance. This is the mathematical expression of the fact that coordinates do not exist {\it a priori} in nature -- being, instead, only artifices used to describe physical systems -- and hence play no role in the formulation of fundamental physical laws. Thus, any fundamental ``quantum" theory of gravity (whatever this may mean) must also be generally covariant. That the gravitational field equations take the form of tensor equations may also be seen as the mathematical expression of the {\it principle of general relativity}. This states that the laws of physics take the same form in all reference systems and is a generalization of the {\it principle of special relativity}, which states that the laws of physics take the same form in all inertial frames. That the gravitational field is equivalent to the curvature of the space-time background follows from applying the principle of general relativity to accelerating frames, combined with local Lorentz invariance \cite{Misner:1974qy}.

The unique, static, spherically symmetric vacuum solution of the EFEs -- which corresponds to the external gravitational field of a point-like particle of mass $M$ -- is the Schwarzschild metric. In canonical polar coordinates this gives the line-element \cite{Dirac2,Misner:1974qy} 
\begin{eqnarray} 
\label{Schwarz_metric}
ds^2 = \left(1 - \frac{2GM}{c^2r}\right)c^2dt^2 - \left(1 - \frac{2GM}{c^2r}\right)^{-1}dr^2 - r^2(d\theta^2 + \sin^2\theta d\phi^2) \, ,
\end{eqnarray}
where $t$ is the time measured by a static observer at $r \rightarrow \infty$. Two key physical properties of the Schwarzschild metric are:
\begin{itemize}

\item{The ``event horizon" at $r = 2GM/c^2$. This is a spherical surface on which the escape velocity is equal to the speed of light $c$. Thus, objects located within the sphere are {\it causally disconnected} from those outside it.}

\item{The singularity at $r=0$. This is a point at which the space-time curvature becomes formally infinite, which marks the break down of the classical theory of general relativity.}

\end{itemize}
The radius of the event horizon in the Schwarzschild solution is called the Schwarzschild radius and is denoted
\begin{eqnarray} 
\label{Schwarz_rad}
r_S = \frac{2GM}{c^2} \, .
\end{eqnarray}


\subsection{The semi-Newtonian approach} \label{Sec.3.2}

Equation (\ref{Newton}) represents Newton's law of gravity for spherically symmetric bodies and results directly from the weak field approximation (\ref{weak_field-1})-(\ref{weak_field-2}) of the Schwarzschild solution (\ref{Schwarz_metric}). As expected, $c$ ``drops out" of the field equations in this limit, yielding a formula which is valid in the non-relativistic gravitational regime. 

Remarkably, we may also ``derive" a crucial relativistic-gravitational result, namely, the existence of the Schwarzschild radius (\ref{Schwarz_rad}), {\it without} imposing the full structure of general relativity. Specifically, we may consider the gravitational field in a Newtonian background of absolute space and time, in which a cosmic speed limit $v_{\rm max} = c$ is imposed ``by hand", {\it without} imposing any of its logical consequences, i.e. local Lorentz invariance and the unification of space and time (in the zero gravity limit), or general covariance and space-time curvature (required in the strong gravity regime).

Combining Newton's laws of motion with Eq. (\ref{Newton}), we obtain the expression for the total energy of a particle of mass $m$, moving in the gravitational field of a body of mass $M$, as
\begin{eqnarray} 
\label{E_tot_Newton}
E_{\rm total} = \frac{1}{2}mv^2 - \frac{GmM}{r} \, .
\end{eqnarray}
The escape velocity, at a given radius $r$, is obtained by setting $E_{\rm total} = 0$, giving 
\begin{eqnarray} 
\label{v_esc}
v_{\rm esc}^2(r) = \frac{GM}{r} \, .
\end{eqnarray}
Setting $v_{\rm esc} \leq c$ then implies that the particle must be situated at a radius 
\begin{eqnarray} 
\label{v_esc}
r \geq r_S = \frac{2GM}{c^2} \, 
\end{eqnarray}
in order to escape from the gravitational field of the massive body. Conversely, test particles initially within $r_S$ remain trapped in the region $r \leq r_S$ for all time. We call this the semi-Newtonian approach, and its advantages for constructing a na{\" i}ve theory of quantum gravity, which permits a unified description of black holes and fundamental particles, are discussed in Sec. \ref{Sec.3.4}. 

\subsection{Newtonian and Einstein gravity -- {\it with} wave-particle duality?} \label{Sec.3.3}

In this sub-section, we review two attempts to incorporate quantum effects into gravitational theories -- one in the Newtonian picture, leading to the Schr{\" o}dinger-Newton equation \cite{Bahrami:2014gwa} -- and the other in background-independent form, via the canonical quantization of general relativity, leading to loop quantum gravity (LQG) \cite{Rovelli:1997yv,Thiemann:2002nj,Smolin:2004sx,Ashtekar:2013hs}.  

\subsubsection{Newtonian gravity -- the Schr{\" o}dinger-Newton equation}  \label{Sec.3.3.1}

One attempt to combine wave-particle duality with the existence of the Newtonian gravitational field is based on the observation that the quantity $m|\psi|^2$ behaves like a ``quantum" mass density for the wave function. Formally identifying this with the source term in Poisson's equation (\ref{Poisson_eq}), we obtain
\begin{eqnarray} 
\label{Poisson_eq_quant}
\vec{\nabla}^2\Phi =  4\pi Gm|\psi|^2 \, .
\end{eqnarray} 
The next step is to replace the external classical potential $V$ in Schr{\" o}dinger's equation (\ref{Schrodinger}) with $m\Phi$, which is supposed to represent the self-gravity field generated by the particle itself. This gives
\begin{eqnarray} 
\label{Schrodinger-Newton}
\left(-\frac{\hbar^2}{2m}\vec{\nabla}^2 + m\Phi \right)\psi = i\hbar\frac{\partial \psi}{\partial t}  \, ,
\end{eqnarray} 
which is known as the Schr{\" o}dinger-Newton equation \cite{Bahrami:2014gwa}. The main problem with this approach is that the explicit solution of Eq. (\ref{Poisson_eq_quant}),
\begin{eqnarray} 
\label{Phi_soln}
\Phi(\vec{x},t) =  -Gm \int \frac{|\psi(\vec{x},t)|^2}{|\vec{x}-\vec{x}'|}d^3x' \, ,
\end{eqnarray}
is {\it semi-classical}, in the sense that it is an explicit function of the classical coordinates $\vec{x}$ $-$ which, here, cannot be interpreted as quantum operators $-$ as well as the wave function $\psi$. Therefore, though useful in models of wave function collapse \cite{Bahrami:2014gwa} and potentially testable in the near future \cite{Grossardt:2015moa,Gan:2015cxz}, Eq. (\ref{Schrodinger-Newton}) cannot be interpreted as the EOM for the wave function of a ``free" quantum particle, interacting only with its own (self-generated) gravitational field. In short, it fails to adequately ``quantize" the particle's self-gravity potential, which remains a function of external classical coordinates. 

In fact, it may be explicitly shown that the Schr{\" o}dinger-Newton equation arises as the weak field limit of semi-classical general relativity, in which only matter fields are quantized, while the gravitational field remains classical even at the fundamental level. This is given by the semi-classical Einstein equations \cite{Bahrami:2014gwa}
\begin{eqnarray} 
\label{Einstein_semi-class-1}
G_{\mu\nu} \equiv R_{\mu\nu} - \frac{1}{2}Rg_{\mu\nu} = \frac{8\pi G}{c^4} \bra{\psi} \hat{T}_{\mu\nu} \ket{\psi} \, ,
\end{eqnarray}
where $\hat{T}^{\mu\nu}$ denotes the set of quantum mechanical operators associated with the components of the classical energy-momentum tensor. In the linearized theory, $g_{\mu\nu} = \eta_{\mu\nu} + h_{\mu\nu}$, where $\eta_{\mu\nu}$ is the Minkowski metric and $h_{\mu\nu}$ represents a small perturbation, Eq. (\ref{Einstein_semi-class-1}) reduces to 
\begin{eqnarray} 
\label{Einstein_semi-class-2}
\vec{\nabla}^2\Phi = \frac{4\pi G}{c^2} \bra{\psi} \hat{T}_{00} \ket{\psi} \, ,
\end{eqnarray}
where $\Phi = -h_{00}c^2/2$. Identifying $ \hat{T}_{00}/c^2 = \hat{\rho} = m|\psi|^2$ in the non-relativistic limit, we recover Eq. (\ref{Schrodinger-Newton}) \cite{Bahrami:2014gwa}. 

\subsubsection{General relativity -- loop quantum gravity}  \label{Sec.3.3.2}

Loop quantum gravity (LQG) represents a canonical quantization of the general theory of relativity; that is, it is based on a map from field variables obeying classical Poisson brackets to commutators involving the corresponding quantum operators. The starting point is, therefore, to rewrite the Einstein-Hilbert action (\ref{Einstein-Hilbert}) in terms of variables satisfying the canonical Poisson bracket structure. These are known as Ashtekar variables \cite{Ashtekar:2015vbm}. The canonical quantization of the gravitational field then proceeds by analogy with Eqs. (\ref{quantization_scheme}) and (\ref{canon_quant_QFT}).

Though, for the sake of brevity, we are unable to give a detailed account of the mathematical formalism of LQG here, we note that, just as the union of wave-particle duality with special relativity gave rise to many unexpected physical phenomena -- including the existence of anti-matter, pair production and quantum mechanical spin -- the canonical quantization of the space-time background (whether or not this leads to final theory of quantum gravity \cite{Carlip:2008zf,Carlip:2001wq}), also yields profound insights and new predictions. Among the most significant are:

\begin{itemize}

\item{That background-independent theories, including general relativity and all related supergravity theories, can be reformulated as gauge theories. In this case, the required gauge symmetry is diffeomorphism invariance, a key property of space-time in general relativity \cite{Ashtekar:2004eh}.}

\item{The physical picture that emerges is of dynamical space-time as a network of interwoven finite ``loops", known as a spin foam \cite{Rovelli:1997yv,Thiemann:2002nj,Smolin:2004sx,Ashtekar:2013hs}.}

\end{itemize}
(The interested reader is referred to \cite{Rovelli:1997yv,Thiemann:2002nj,Smolin:2004sx,Ashtekar:2013hs} and references therein for a full summary of the present state of LQG research.) Strictly, in this picture, the ``quantum" nature of gravity arises from the fact that space-time can exist as a superposition of spin foams. These represent generalized Feynman diagrams \cite{Peskin:1995ev} which, instead of graphs, use higher-dimensional complexes. A spin foam represents the time evolution of spin network -- a specific type of $2$-complex whose faces represent possible configurations of the field -- that must be summed to obtain an amplitude via the path integral approach \cite{Rovelli:1997yv,Thiemann:2002nj,Smolin:2004sx,Ashtekar:2013hs}.

\subsection{The semi-Newtonian approach revisited} \label{Sec.3.4}

In Sec. \ref{Sec.3.2}, we saw that the imposition of a cosmic speed limit, $v_{\rm max} = c$, in a Newtonian background of absolute space and time, allowed us to recover the expression for the Schwarzschild radius (\ref{Schwarz_rad}) from the formula for the escape velocity. This may be considered a fundamental result of ``semi-Newtonian" gravity, that, happily, recovers the correct general-relativistic formula. In Sec. \ref{Sec.2.1}, a limit on the momentum uncertainty of single quantum particle, in the non-relativistic theory, $(\Delta_{\psi}p)_{\rm max} \approx mc$, was obtained by noting that, for $(\Delta_{\psi}x)_{\rm min} \approx \lambda_C$, relativistic field effects imply pair-production instead of further localization of the wave packet. We now note that the identifications in Eq. (\ref{Compton-HUP}) may be obtained in an alternative way, via the imposition of the semi-Newtonian condition in canonical QM. Beginning with the non-relativistic formula for the momentum of a point-particle, 
\begin{eqnarray} 
\label{p=mv}
\vec{p} = m\vec{v} \, ,
\end{eqnarray}
and imposing the cosmic speed limit, gives $p \leq mc$. Using $\Delta_{\psi}p \lesssim p$, we then have $(\Delta_{\psi}p)_{\rm max} \approx mc$, so that the HUP (\ref{HUP}) implies $(\Delta_{\psi}x)_{\rm min} \approx \lambda_C$. 

The logic used here is the reverse of our previous argument in Sec. \ref{Sec.3.2}, in which a relativistic result (the existence of $\lambda_C$) was used to impose a limit on the spatial extent of a non-relativistic wave packet. This, combined with the HUP (\ref{HUP}), then gave rise to a limit on the spread in momentum space. In the semi-Newtonian approach, we use a result from the non-relativistic theory (\ref{p=mv}), together with $v_{\rm max} = c$, to obtain a limit on the momentum uncertainty, which then implies the existence of $\lambda_C$. This should come as no surprise; since the Compton wavelength marks the boundary between relativistic and non-relativistic quantum theory, approaching the line $\lambda = \lambda_C$ from either side should give the same results. 

However, the realization that the Compton wavelength (\ref{Compton}) may be obtained by applying the semi-Newtonian condition to the general structure of canonical QM (which ``lives" in a Newtonian background) {\it and} that the same condition, applied to Newtonian gravity, gives rise to the Schwarzschild radius (\ref{Schwarz_rad}), is important. In principle, it allows us to construct a ``semi-Newtonian" theory of quantum gravity, in which essential phenomenological features of both relativistic QM and relativistic gravity are recovered in the appropriate asymptotic regimes. Such a theory would be na{\" i}ve, failing even to account correctly for the local Lorentz invariance of special relativity, let alone general covariance. Nonetheless, it could be interesting, giving rise to a quantum mechanical description of massive objects which is consistent with key general-relativistic results, at least for spherically symmetric systems, as well as key results from relativistic QFT. It would, necessarily, imply a unified description of fundamental particles and black holes, as quantum objects, in an absolute background with a cosmic speed limit. 

With this in mind, we note that $\lambda_C$ and $r_S$ define radii with fundamentally {\it different} physical natures; $\lambda_C$ defines a minimum radius, beyond which further compression is impossible, whereas $r_S$ defines a maximum radius, beyond which further compression is inevitable. Since any unified picture of fundamental particles and black holes must reflect this, the question arises, what form of wave-particle duality defines their (unified) ``quantum" nature? Does it take the same form in each asymptotic limit? The different physical natures of $\lambda_C$ and $r_S$ would suggest ``no". It is likely, therefore, that a unified theory could be based on a extension of the notion of wave-particle duality, which reduces to the canonical de Broglie relations (\ref{deBroglie}) for small mass objects (particles), but takes a different asymptotic form for large mass objects (black holes).  

Hence, for our present purposes, in which we seek only a na{\" i}ve, semi-Newtonian, theory of quantum gravity, the most relevant insight of LQG, discussed in Sec. \ref{Sec.3.3.2}, is the following: Since the Ashtekar variables necessarily involve both $G$ and $c$, and the canonical map $\lim_{\hbar \rightarrow 0}\left\{ \ , \ \right\}/(i\hbar) = [ \ , \ ]$ relates these to Planck's constant $\hbar$, it is reasonable to assume that, in the semi-Newtonian regime, this gives rise to an effective dispersion relation for the gravitational field involving $G$, $c$ and $\hbar$. More specifically, we expect a modification of the standard dispersion relation for a (non-gravitating) Newtonian particle, containing all three constants, which reduces to Eq. (\ref{non_rel_disp}) in the appropriate limit. In principle, this may be achieved in two ways: 

\begin{itemize}

\item{By substitution of the canonical de Broglie relations (\ref{deBroglie}) into a modified, semi-classical, formula for the {\it total} energy of a self-gravitating particle (Eq. (\ref{non_rel_E-p}) with $V \equiv V(\psi) \neq 0$).}

\item{By substituting modified de Broglie relations into the existing Newtonian formula for a {\it free} particle (Eq. (\ref{non_rel_E-p}) with $V=0$).}

\end{itemize}
The first approach results, most naturally, in the Schr{\" o}dinger-Newton equation (\ref{Schrodinger-Newton}), and does not allow us to accomplish our desired aim, as discussed in Sec. \ref{Sec.3.3.1}. We therefore adopt the second. However, since, mathematically, we may consider arbitrary modifications of Eq. (\ref{deBroglie}), we adopt the following criteria as guidelines for a physical theory:

\begin{itemize}

\item{We require the modified dispersion relation to be consistent with essential phenomenological features of both relativistic QM and relativistic gravity -- specifically, the existence of the Compton wavelength for fundamental particles and the Schwarzschild radius for black holes -- in appropriate limits.}

\item{Within the constraints imposed by the asymptotic regimes, the new expression should be as simple as possible, containing the minimum number of new fundamental parameters.}

\item{All physically observable quantities must be continuous, including at the transition point between quantum particles and quantum black holes (i.e., close to the Planck point).}

\item{The basic mathematical structure of canonical QM, and its relation to measurements of physical observables, must be preserved. This implies that, of the four postulates \cite{Ish95} of the canonical theory, only Postulate $4$, which states that the time-evolution of the quantum state $\ket{\psi}$ is given by the Schr{\" o}dinger equation (\ref{Schrodinger}), must be revised. }

\item{In particular, the time-evolution given by the new EOM, corresponding to the modified dispersion relation, must be unitary.}

\end{itemize}

The first criteria implies that dependence on $G$ and $c$ should drop out of the modified dispersion relation when $m$ is small, whereas $\hbar$ should drop out when $m$ is large, relative to the Planck mass. As we shall see in Sec. \ref{Sec.4}, this requires ``extending" the notion of wave-particle duality into the mass range in which the canonical relations (\ref{deBroglie}) break down, namely, the black hole region of the mass-radius diagram. We also note that, though modified dispersion relations necessarily imply modified EOM, this is simply equivalent to modified expressions for the Hamiltonian and momentum operators. Specifically, the Hilbert space structure of the theory and, hence, its probabilistic interpretation, are {\it unaffected} by this change, as long as the time-evolution of the state vector $\ket{\psi}$ remains unitary.

\section{Extended de Broglie relations} \label{Sec.4}

The Planck mass and length scales are obtained via dimensional analysis using the fundamental constants $G$, $c$ and $\hbar$:
\begin{eqnarray} 
\label{Planck}
m_P = \sqrt{\frac{\hbar c}{G}} = 2.177 \times 10^{-5}  {\rm g} \, , \quad \quad l _P = \sqrt{\frac{\hbar G}{c^3}} = 1.616 \times 10^{-33} {\rm cm} \, .
\end{eqnarray}
(Strictly, these are the reduced Planck scales, but we will refer to them simply as the Planck scales from here on.) $l_P$ is believed to be the shortest resolvable distance due to quantum gravity effects, corresponding to irremovable quantum fluctuations in the metric \cite{Padmanabhan:1986ny}. It is therefore unclear what (if any) meaning can be attributed to de Broglie waves with $\lambda \lesssim l_P$. In fact, the non-relativistic energy-momentum relation (\ref{non_rel_E-p}) implies that the angular frequency and wave number of the de Broglie waves reach the Planck values
\begin{eqnarray} \label{Planck*}
\omega_P = \frac{2\pi}{t_P} \, ,  \quad \quad  k_P =\frac{ 2\pi}{l_P} \, ,
\end{eqnarray}
where $t_P = l_P/c$ is the Planck time, for free particles with energy, momentum and mass given by $E = pc = 2\pi m_Pc^2$ and $m = \pi m_P$, respectively. Since a further increase in energy would imply a de Broglie wavelength smaller than the Planck length, it is unclear how quantum particles behave for $E \gtrsim m_Pc^2$. 

However, ignoring numerical factors of order unity, both $m_P$ and $l_P$ also admit intuitive physical interpretations as the mass and radius, respectively, of an object that is simultaneously a particle {\it and} a black hole. In other words, for a ``particle" of mass $m \approx m_P$, the standard formulae for $\lambda_C$ and $r_S$ give $\lambda_C \approx r_S \approx l_P$. We note that this involves extrapolating the standard results of relativistic (non-gravitational) quantum theory and relativistic (non-quantum) gravity to their extreme limit. In the standard scenario, a log-log plot of mass vs radius -- hereafter referred to simply as the $(M,R)$ diagram -- is separated into two disjoint halves, with the Compton radius of quantum-relativistic (non-gravitational) particles on the left ($m \lesssim m_P$) and the Schwarzschild radius of relativistic-gravitational (non-quantum) black holes on the right ($m \gtrsim m_P$). In this plot, the two lines are symmetric under reflection in the line $m = m_P'$ (where $m_P' \approx m_P$), which corresponds to the transformation $m \leftrightarrow m_P'^2/m$. (This symmetry is discussed further in Sec. \ref{Sec.4.2}.) But, since both the standard formulae for $\lambda_C$ and $r_S$ may be recovered in the semi-Newtonian approximation, this raises the intriguing possibility that a modified form of canonical QM, based on a dispersion relation involving $G$, $c$ and $\hbar$, could yield a unified description of particles and black holes that unites $\lambda_C$ and $r_S$ into a single line. In this scenario, ``particles" with $E \gtrsim m_Pc^2$ become black holes, and the modified dispersion relation is required to prevent $\lambda \lesssim l_P$ in this region of the $(M,R)$ diagram.

Though it may be argued that the Newtonian formula is not valid close to the Planck scales, due to both relativistic and gravitational effects, this is not necessarily the case. In non-gravitational theories, relativistic effects are important for particles whose kinetic energy higher is than their rest mass energy, $E_{\rm kinetic} \gtrsim mc^2$. However, a particle (i.e. a small, spherically symmetric body) can have Planck scale energy if it has Planck scale rest mass, even if it is moving non-relativistically. In the following analysis, we consider the rest frame of quantum particles with rest masses extending into the regime $m \gg m_P$, for which the the standard non-relativistic analysis is expected to hold, except for modifications induced by gravity (rather than relativistic velocities). This is equivalent to applying the semi-Newtonian approximation on both the left-hand and right-hand sides of the $(M,R)$ diagram.

Figure 1 shows the Compton and Schwarzschild lines -- here denoted $\lambda_C$ and $\lambda_S$ for notational consistency -- as asymptotes in the $m \ll m_P$ and $m \gg m_P$ regimes, respectively. The line $\lambda_{C/S}'$ represents a possible interpolation between the two regions, in which $\lambda_C$ and $\lambda_S$ are unified into a single curve. Since the Compton wavelength may also be ``derived" from the HUP in canonical QM by applying the semi-Newtonian approximation, as discussed in Sec. \ref{Sec.3.4}, a unification of this form is known as the black hole uncertainty principle (BHUP) correspondence \cite{Ca:2014,cmp}, as well as the Comtpon-Schwarzschild  correspondence \cite{Lake:2015pma}. Crucially, we again note that any would-be unified description of particles and black holes must account for the change in the physical nature of the radius $\lambda_{C/S}'$ at $m \approx m_P$, from minimum to maximum spatial extent of the wave packet. This is depicted visually in Fig. 2. In the language of semi-Newtonian QM, the shaded region on the left-hand side of Fig. 2 is associated with the ``$\geq$" inequality from the HUP. It is therefore reasonable to expect that the shaded region on the right should be associated with the inequality ``$\leq$" in a modified uncertainty relation, giving rise to a different kind of positional uncertainty for black holes. The minimum of the curve $\lambda_{C/S}'$ is then associated with equality ``$=$" in which the modified Compton and Schwarzschild radii are equal. This state is unique, with $(\Delta_{\psi}x)_{\rm min} = (\Delta_{\psi}x)_{\rm max} \approx l_P$. (Both Fig. 1 and Fig. 2 are taken from \cite{Lake:2015pma}.)

Finally, we note that, in the semi-Newtonian picture, the existence of a minimum radius for the wave packet $(\Delta_{\psi}x)_{\rm min}$ for $m \lesssim m_P$ implies a UV cut-off for $\vec{k}$ in the usual Fourier expansion of $\ket{\psi}$. Likewise, the existence of a maximum radius $(\Delta_{\psi}x)_{\rm max}$ for $m \gtrsim m_P$ implies an IR cut-off in $\vec{k}$ for the wave functions black holes. (That these cut-offs are not fundamental, but instead mark the point at which the semi-Newtonian approximation breaks down, and relativistic effects must be more fully accounted for, need not concern us.) Therefore, if a unified description can be  based on a modified quantum dispersion relation, we expect such limits to arise naturally from the relations themselves. 

\begin{figure}[h] \label{Fig.1}
\centering
\includegraphics[width=12cm]{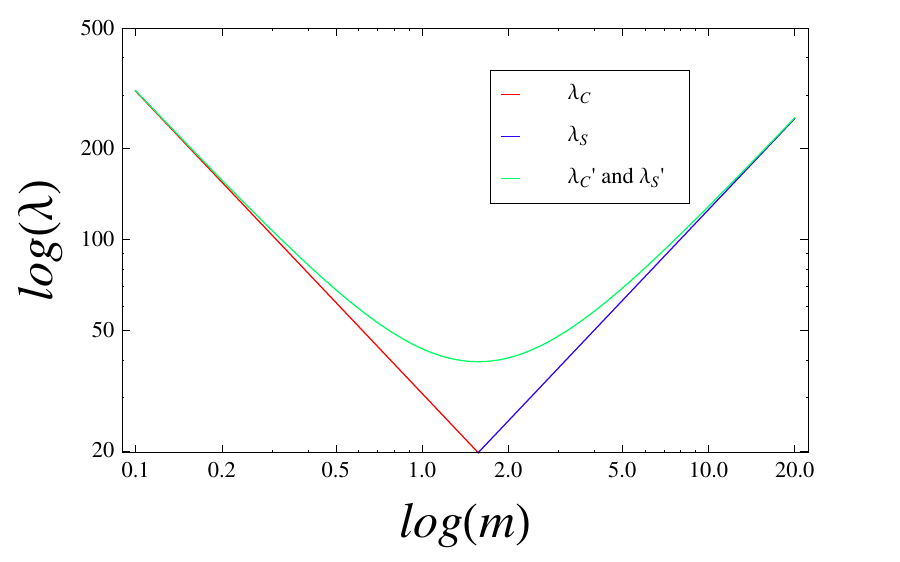}
\caption{The Compton wavelength $\lambda_{C} \propto 1/m$ (red line) and Schwarzschild radius $\lambda_{S} \propto m$ (blue line) as asymptotes to a unified radius $\lambda'_{C/S}$ (green curve) for objects with any rest mass \cite{Lake:2015pma}.}
\end{figure}

\begin{figure}[h] \label{Fig.2}
\centering
\includegraphics[width=11cm]{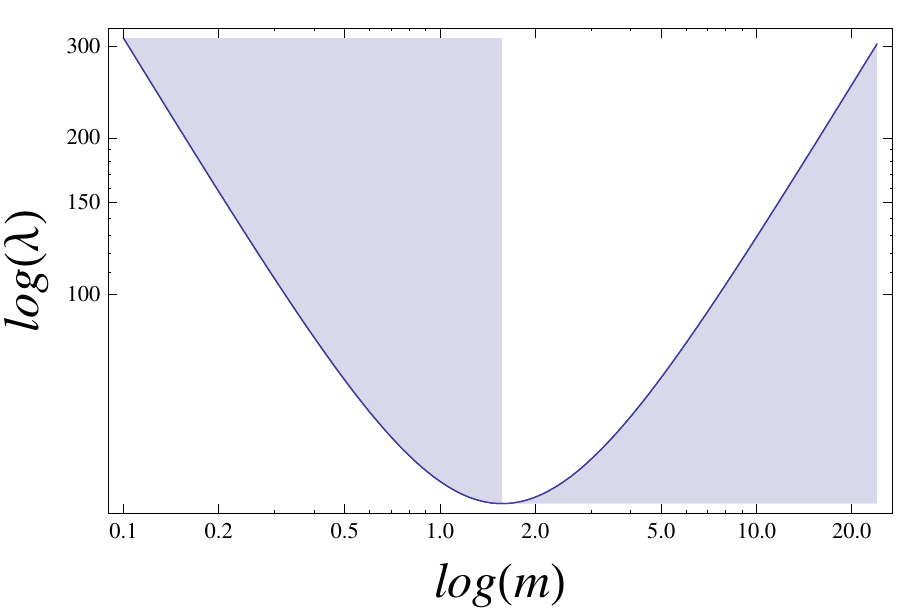}
\caption{The physical nature of the radius $\lambda'_{C/S}$ changes dramatically at the minimum close to $m=m_P$. Particles are defined by the existence of a minimum radius; attempts to compress the width of the wave packet beyond this result in pair-production, rather than further spatial localization. Black holes are defined by the existence of a maximum radius; within this, the wave packet associated with the central black hole mass is ``contained", indicating the causal disconnect between the black hole's interior and its exterior surroundings \cite{Lake:2015pma}. (Nonetheless, in principle, this picture remains consistent with the existence of Hawking radiation, which may be interpreted as the result of QM tunneling through an effective potential boundary, located in the region of the classical black hole horizon.)}
\end{figure}

\subsection{de Broglie relations for black holes -- the limit $E \gg m_Pc^2$} \label{Sec.4.1}

An arbitrary generalization of the usual (one-dimensional) de Broglie relations may be written as
\begin{eqnarray} 
\label{deBroglie_ext}
E = \hbar\Omega(\omega) \, , \quad \quad p = \hbar\kappa(k) \, .
\end{eqnarray}
where $\Omega(\omega)$ and $\kappa(k)$ are arbitrary functions of the angular frequency and wave number, respectively. (For the sake of simplicity, we restrict ourselves to one-dimension, which may also be interpreted as an analysis of spherically symmetric systems.) Let us now suppose that, in the limit $E \gg m_Pc^2$, the generalized de Broglie relations (\ref{deBroglie_ext}) take the form 
\begin{eqnarray} 
\label{deBroglie_m>>m_P}
E \approx \hbar\beta\frac{\omega_P^2}{\omega} \, , \quad \quad p \approx \hbar \beta \frac{k_P^2}{k} \, ,
\end{eqnarray} 
where $\beta > 0$ is a dimensionless constant. Combining (\ref{non_rel_E-p}) (with $V=0$) and (\ref{deBroglie_m>>m_P}), gives the dispersion relation
\begin{eqnarray} 
\label{disp_rel_m>>m_P*}
\omega \approx \frac{2 G m}{(2\pi)^2 \beta c}k^2 \, .
\end{eqnarray}
Assuming that the momentum operator eigenfunctions take the usual form (\ref{posn_space_eigenfn}), the EOM for the quantum state is of Schr{\" o}dinger type:
\begin{eqnarray} 
\label{Schrod_1D}
-\frac{mc^2}{k_P^2}\frac{\partial^2 \psi}{\partial x^2} \approx \frac{i\beta\hbar}{2} \frac{\partial \psi}{\partial t} \, ,
\end{eqnarray}
which corresponds to modified Hamilton and momentum operators, $\hat{H}'$ and $\hat{p}'$. The differential part of $\hat{p}'$ takes the standard form, but the multiplying factor differs from the canonical one; the first corresponds to the functional form of the dispersion relation, while the second sets the phenomenologically important length scale for the $m \gg m_P$ regime. We note that this is proportional to $m$ and is of the same order of magnitude as $\lambda_S$, whereas the canonical de Broglie relations yield a natural length scale proportional to $1/m$, of the order of $\lambda_C$. Assuming that the position operator may be defined as usual, we have
\begin{eqnarray}
\label{p_x_1D}
\hat{p}' \approx -i\frac{\sqrt{2}mc}{k_P}\frac{\partial}{\partial x} \, , \quad \quad  \hat{x} = x \, ,
\end{eqnarray}
so that the commutator is
\begin{eqnarray} \label{x-p_comm_m>>m_P}
[\hat{x},\hat{p}'] = i\frac{\sqrt{2}mc}{k_P} \, .
\end{eqnarray}

The proposed asymptotic form of the extended de Broglie relations (\ref{deBroglie_m>>m_P}) disturbs the mathematical structure of canonical QM only so far as the momentum operator is multiplied by a constant factor. This modifies the Hamiltonian and momentum operators (and functions of these), together with the commutation relations, but does not affect the underlying Hilbert space structure of the theory. Postulate 4 is changed, but only trivially \cite{Ish95}. The underlying mathematical formalism is the same as in canonical QM, so that the new theory is well defined in the limit $E \gg m_Pc^2$. 
 
The physical motivation for suggesting the asymptotic form (\ref{deBroglie_m>>m_P}) is the natural emergence of the Schwarzschild radius as a phenomenologically significant length scale. The modified uncertainty relation gives
\begin{eqnarray} 
\label{x-p_SUP_m>>m_P}
\Delta_{\psi}x\Delta_{\psi}p' \geq \frac{1}{2}\braket{\psi| [\hat{x},\hat{p}'] |\psi} \approx \frac{m \, c \, l_P}{2\sqrt{2}\pi} \, .
\end{eqnarray}
For $m \ll m_P$, we use (\ref{HUP}), together with the semi-Newtonian approximation, to make the identifications (\ref{Compton-HUP}). By contrast, using (\ref{x-p_SUP_m>>m_P}), together with the semi-Newtonian approximation for $m \gg m_P$, gives
\begin{eqnarray} 
\label{x_max_m>>m_P}
(\Delta_{\psi}x)_{\rm max} \approx \frac{l_P^2}{\lambda_C} \approx \lambda_S \, , \quad \quad (\Delta_{\psi}p')_{\rm min} \approx mc \, .
\end{eqnarray} 
Substituting (\ref{x_max_m>>m_P}) into (\ref{x-p_SUP_m>>m_P}), and ignoring numerical factors of order unity, gives $m \gtrsim m_P$, in accordance with our original assumption about the validity of Eq. (\ref{x-p_SUP_m>>m_P}) in this range. Thus, the identifications (\ref{x_max_m>>m_P}) are consistent with the assumption $E \gtrsim m_Pc^2$.

\subsection{de Broglie relations for all energies} \label{Sec.4.2}

We now propose a physically viable ansatz for the theory of extended de Broglie relations, which is able to provide a unified description of particles and back holes in the semi-Newtonian approximation. This is intended as a toy model, since deviations from canonical QM behaviour due to gravitational effects have yet to be detected experimentally \cite{Grossardt:2015moa,Gan:2015cxz}, and current observations offer no clue regarding the precise mathematical form of the quantum dispersion relation close to the Planck point. In principle, any curve $\lambda_{C/S}'$ that asymptotes to both $\lambda_{C}$ and $\lambda_{S}$ in the appropriate regimes remains phenomenologically viable at the present time. 

In this section, we show that, for de Broglie relations that hold for all energies, and take appropriate asymptotic forms for $E \ll m_Pc^2$  and $E \gg m_Pc^2$, the quantum EOM is \emph{not}, in general, of Schr{\" o}dinger type. Deviations from Schr{\" o}dinger-type evolution are most pronounced for $m \approx m_P$, though the EOM simplify to (\ref{Schrodinger}) and (\ref{Schrod_1D}) in the small and large mass limits, respectively. Nonetheless, both unitarity and the Hilbert space structure of the canonical theory are preserved throughout the entire mass range. Once again, only Postulate 4 of \cite{Ish95} must be amended, this time less trivially, while the mathematical structure embodied in Postulates 1, 2 and 3 is unchanged.

The ``simplest" form of the functions $\Omega(\omega)$ and $\kappa(k)$ in Eq. (\ref{deBroglie_ext}), able to satisfy the physical requirements outlined in Sec. \ref{Sec.3.4}, are:
\begin{equation} 
\label{Omega1}
\Omega(\omega) = \left \lbrace
\begin{array}{rl}
\omega_P^2\left(\omega + \omega_P^2/\omega\right)^{-1} & \ (m \lesssim m_P) \\
\beta\left(\omega + \omega_P^2/\omega\right) & \ (m \gtrsim m_P) \ \ \
\end{array}\right.
\end{equation}
\begin{equation} 
\label{kappa1}
\kappa(k) = \left \lbrace
\begin{array}{rl}
k_P^2\left(k + k_P^2/k\right)^{-1} & \ (m \lesssim m_P) \\
\beta\left(k + k_P^2/k\right) & \ (m \gtrsim m_P).
\end{array}\right.
\end{equation}
Though ``simplicity" remains a subjective judgement, here we take it to mean that the new functions contain only terms in $\omega$ and $\omega^{-1}$ or $k$ and $k^{-1}$, in the $m \gg m_P$ limit, reflecting the $m \leftrightarrow 1/m$ symmetry of the $(M,R)$ diagram, with only the minimum number of additional free parameters -- in this case, the single numerical constant $\beta > 0$. We note that the continuity of $E$, $p$, $dE/d\omega$ and $dp/dk$, at $\omega = \omega_P$, $k=k_P$, implies $\beta = 1/4$. However, since we require continuity of {\it all} physical observables at $m \approx m_P$, and $\beta$ must be fixed to ensure this, we leave it as a free parameter for now. 

For $m \lesssim m_P$, combining Eqs. (\ref{Omega1})-(\ref{kappa1}) with (\ref{non_rel_disp}), gives  
\begin{eqnarray} \label{disp_rel_m<m_P1**}
\omega^2 - 2ck_C\left(\frac{k}{k_P} + \frac{k_p}{k} \right)^2 \omega + \omega_P^2 = 0 \, ,
\end{eqnarray} 
which reduces to $\omega \approx (\hbar/2m)k^2$ for $\omega \ll \omega_P$ and $k \ll k_P$, as required. In general, however, (\ref{disp_rel_m<m_P1**}) may be solved to give {\it two} solution branches,
\begin{eqnarray} \label{disp_rel_m<m_P1}
\omega_{\pm}(k:m) = ck_C\left(\frac{k}{k_P} + \frac{k_p}{k} \right)^2\left[1 \pm \sqrt{1 - \frac{k_P^2}{k_C^2}\left(\frac{k}{k_P} + \frac{k_P}{k}\right)^{-4}} \right] \, .
\end{eqnarray} 
The reality of both branches requires
\begin{eqnarray} \label{disp_rel_m<m_P3}
k^4 + 2\left(1 - \frac{k_p}{2k_C} \right)k_P^2k^2 + k_P^4 \geq 0 \, ,
\end{eqnarray} 
and the inequality is saturated for 
\begin{eqnarray} \label{disp_rel_m<m_P4}
k_{\pm}^2(m) = k_P^2\left(\frac{k_P}{2k_C} - 1\right)\left[1 \pm \sqrt{1 - \left(\frac{k_P}{2k_C} - 1\right)^{-2}} \right] \, .
\end{eqnarray} 
The reality of these limits then implies
\begin{eqnarray} \label{disp_rel_m<m_P5}
m \leq m_P' \equiv (\pi/2)m_P \, .
\end{eqnarray} 
The inequality (\ref{disp_rel_m<m_P3}) is satisfied for two {\it disjoint} ranges:
\begin{eqnarray} \label{disjoint_k}
k^2 \leq k_{-}^2(m) \ [-k_{-}(m) \leq k \leq +k_{-}(m)] \, , \quad k^2 \geq k_{+}^2(m) \ [k \leq - k_{+}(m) \ {\rm or} \ k \geq +k_{+}(m)] \, . 
\end{eqnarray} 
Defining $k_{\pm}(m) = 2\pi/\lambda_{\mp}(m)$, the lower limit on the value of the wavenumber $k_{-}(m)$ corresponds to an upper limit on the wavelength $\lambda_{+}(m)$ while the upper limit $k_{+}(m)$ corresponds to a lower limit $\lambda_{-}(m)$, so that (\ref{disjoint_k}) is equivalent to 
\begin{eqnarray} \label{disjoint_lambda}
\lambda^2 \geq \lambda_{+}^2(m) \  [\lambda \leq - \lambda_{+}(m) \ {\rm or} \ \lambda \geq +\lambda_{+}(m)] \, , \quad \lambda^2 \leq \lambda_{-}^2(m) \  [-\lambda_{-}(m) \leq \lambda \leq +\lambda_{-}(m)].
\end{eqnarray} 
For $m \ll m_P'$, Eq. (\ref{disp_rel_m<m_P4}) can be expanded to first order, giving
\begin{eqnarray} \label{kapprox}
k_{-}(m) \approx \sqrt{k_Ck_P} \, , \quad \quad k_{+}(m) \approx \sqrt{k_Sk_P} \, ,
\end{eqnarray} 
where $k_S \equiv 2\pi/\lambda_S$, which is equivalent to 
\begin{eqnarray} \label{lambdalimit}
\lambda_{+}(m) \approx \sqrt{\lambda_C l_P}  \, , \quad \quad \lambda_{-}(m) \approx \sqrt{\lambda_S l_P} \, .
\end{eqnarray} 
We note that the first range of wavelengths given by Eqs. (\ref{disjoint_k})-(\ref{disjoint_lambda}) is compatible with the existence of an ``outer" radius, which acts as a minimum {\it obervable} width for the particle, whereas the second range is compatible with the existence of an ``inner" radius, within which sub-Planckian de Broglie modes become trapped, but which remains {\it unobservable}. Though it is unclear whether this range is physically meaningful, we note that, formally extending the $(M,R)$ diagram to $R < l_P$ implies $\lambda_C > l_P$ and $\lambda_S < l_P$ for $m \lesssim m_P$. In this region, $\lambda_S$ acts like an (inner) unobservable radius, which is disjoint from the (outer) observable radius, $\lambda_C$. 

Interestingly, since $\Delta_{\psi}x \gtrsim \lambda_C \geq \lambda_{+}(m)$, Eq. (\ref{lambdalimit}) gives rise to a minimum length uncertainty relation (MLUR) closely resembling the one originally derived, using heuristic arguments, by Salecker and Wigner \cite{Salecker:1957be} (see also \cite{AmelinoCamelia:1999ni} for recent developments) and later considered by Ng and Vam Dam \cite{Ng:1994zk}: $(\Delta x)_{\rm min} \approx \sqrt{\lambda_C d}$, where $d$ is the distance ``probed" using a QM measuring device \cite{Hossenfelder:2012jw}. For $d \geq l_P$, the predictions of the extended de Broglie theory coincide with this relation. However, note that, here, we use the notation $\Delta x$, rather than $\Delta_{\psi} x$, to indicate that the ``uncertainty" obtained in \cite{Salecker:1957be} differs from the canonical definition given in Eq. (\ref{SUP}). In summary, for $m \ll m_P$, the extended de Broglie theory implies $(\Delta_{\psi}x)_{\rm min} \approx \lambda_C$ and  $(\Delta x)_{\rm min} \approx \lambda_{+}(m) \approx \sqrt{\lambda_Cl_P}$, where $(\Delta_{\psi}x)_{\rm min}$ and $(\Delta x)_{\rm min}$ differ as described in \cite{Salecker:1957be,AmelinoCamelia:1999ni,Ng:1994zk,Hossenfelder:2012jw}.

For $m \gtrsim m_P$, the counterparts of Eqs. (\ref{disp_rel_m<m_P1**})-(\ref{disp_rel_m<m_P4}) may be obtained by performing the substitution
\begin{eqnarray} \label{mdual}
m \rightarrow  \frac{\beta\pi^2m_P^2}{m} \, ,
\end{eqnarray} 
and the reality conditions on $k^2_{\pm}(m)$ yield
\begin{eqnarray} \label{disp_rel_m>m_P5}
m \geq 2\pi \beta m_P \, .
\end{eqnarray}
Again, two discontinuous wavelength ranges are permitted, according to Eq. (\ref{disjoint_lambda}), but the new limits are given by
\begin{eqnarray} \label{lambdalimit2}
\lambda_{+}(m) \approx \sqrt{\lambda_S l_P}  \, , \quad \quad \lambda_{-}(m) \approx \sqrt{\lambda_C l_P} \, .
\end{eqnarray}
These are equivalent to those given in Eq.~\eqref{lambdalimit} under the interchange $\lambda_C \leftrightarrow \lambda_S$, up to numerical factors of order unity. This is compatible with the existence of an observable outer radius $\lambda_S > l_P$ and an unobservable inner radius $\lambda_C < l_P$. 

From Eqs. (\ref{disp_rel_m<m_P5}) and (\ref{disp_rel_m>m_P5}), we see that $\beta = 1/4$ is required to ensure the continuity of physical quantities at close to the Planck point. We then have
\begin{equation} 
\label{Omega2}
\Omega(\omega) = \left \lbrace
\begin{array}{rl}
\omega_P^2\left(\omega + \omega_P^2/\omega\right)^{-1} & \  (m \le m_P') \\ (1/4)\left(\omega + \omega_P^2/\omega\right) & \ (m \geq m_P')
\end{array}\right.
\end{equation}
\begin{equation} 
\label{kappa2}
\kappa(k) = \left \lbrace
\begin{array}{rl}
k_P^2\left(k + k_P^2/k\right)^{-1} & \ (m \leq m_P') \\
(1/4)\left(k + k_P^2/k\right) & \ (m \geq m_P') \, ,
\end{array}\right.
\end{equation}
where $m_P'$ is defined in Eq. (\ref{disp_rel_m>m_P5}), giving  $\omega = \omega_P$, $k=k_P$ and $E = pc = 2m_P' c^2$ for $m = m_P'$. The continuity of $k_{\pm}^2(m)$ at $k_P$ and $\omega_{\pm}(k,m)$ at $\omega_P$ is ensured and the reality conditions justify our initial assumptions about the division between $m \lesssim m_P$ and $m \gtrsim m_P$. Note that the original, approximate, conditions given in Eq. (\ref{Omega1}) have now been replaced by the more precise expressions $m \leq m_P'$ and $m \geq m_P'$ in Eq. (\ref{Omega2}).

The solutions obtained above exhibit various dualities: Each branch of the quantum dispersion relation for particles (now defined as spherically symmetric objects with rest masses $m \leq m_P'$) maps to its equivalent for black holes (with rest masses $m \geq m_P'$), under   
\begin{equation} 
\label{dual}
m \leftrightarrow \frac{m_P'^2}{m} \, .
\end{equation}
This is equivalent to a reflection about the line $m = m_P'$ in the log-log plot of the $(\omega(k:m), m)$ plane, at fixed $k$. For fixed $m$, the self-duality
\begin{equation} 
\label{self-dual_omega}
\omega_{\pm} \leftrightarrow \frac{\omega_P^2}{\omega_{\mp}} \, , 
\end{equation}
maps the upper and lower branches of the solution $\omega_{\pm}(k:m)$ into one another in the $(\omega(k:m), k)$ plane, and 
\begin{equation} 
\label{self-dual_k}
k_{\pm} \leftrightarrow \frac{k_P^2}{k_{\mp}} \, 
\end{equation}
transforms the upper and lower limits on the wave number into one another. 

Using the standard definitions of $\lambda_C$ and $\lambda_S$, Eq. (\ref{dual}) also induces the transformation $k_C \leftrightarrow (\pi/8)(k_P^2/k_C) = (\pi/4)k_S$, which is equivalent to $\lambda_S \leftrightarrow (\pi/4)\lambda_C$. However, since $\lambda_C$ gives the order of magnitude length scale at which pair production rates become significant, the exact numerical factor used to define it is to some degree arbitrary, whereas the definition of $\lambda_S$ comes from an exact solution of the EFEs. We are therefore free to redefine $\lambda_C$, such that 
\begin{equation} 
\label{Compton_redef}
\lambda_C \equiv \frac{\pi}{4}\frac{\hbar}{mc} \, .
\end{equation}
The new definition gives
\begin{equation} 
\label{}
\lambda_C =  \frac{l_P^2}{\lambda_S} \, ,
\end{equation}
so that the duality transformation (\ref{dual}) induces
\begin{equation} 
\label{}
\lambda_{C/S} \leftrightarrow \lambda_{S/C}  \, .
\end{equation}
The new definition (\ref{Compton_redef}) also ensures $\lambda_C = \lambda_S = l_P$ at $m = m_P'$, so that we may directly associate the divide between $m \leq m_P'$ and $m \geq m_P'$ in Eqs. (\ref{Omega2})-(\ref{kappa2}) with the transition between particles and black holes. For either definition, (\ref{Compton}) or (\ref{Compton_redef}), the equivalents of Eqs. (\ref{disp_rel_m<m_P1**})-(\ref{disp_rel_m<m_P4}) for the black hole regime may be obtained (up to factors of order unity multiplying the relevant terms), via the substitution $k_C \leftrightarrow k_S$. The asymptotic regimes of $\omega_{-}(k:m)$ are then given by 
\begin{equation} 
\label{}
\omega_{-}(k:m) \approx \frac{1}{2}\left(\frac{k}{\sqrt{k_{S/C}k_p}} + \frac{\sqrt{k_{S/C}k_p}}{k}\right)^{-2}ck_P \, ,
\end{equation}
where we choose $k_S$ for $m \leq m_P'$ and $k_C$ for $m \geq m_P'$. For weakly gravitating particles, $m \ll m_P'$ and $|k| \leq \sqrt{k_Ck_P} \ll k_P \ll \sqrt{k_Sk_P}$, whereas, for black holes, $m \gg m_P'$ and $|k| \leq \sqrt{k_Sk_P} \ll k_P \ll \sqrt{k_Ck_P}$. Both give $\omega_{-}(k:m) \propto k^2 \ll \omega_P^2$.   

To summarize: Formally, there exist two {\it inequivalent} dispersion relations, in each mass range $m \leq m_P'$ and $m \geq m_P'$, corresponding to the separate branches $\omega_{\pm}(k:m)$, though the physical interpretation of the $\omega_{+}(k:m)$ branch is unclear, as this corresponds to de Broglie waves with sub-Planckian wavelengths. The super-Planckian branch $\omega_{-}(k:m)$ depends on both $k$ and $m$, as in canonical QM, but, in the modified theory, there exist limiting values of $k$, which are also functions of the rest mass. This may be contrasted with the canonical non-relativistic theory, in which all modes $k \in (-\infty,\infty)$ may contribute, with some nonzero amplitude, to a given wave packet expansion. 

All four solution branches, $\omega_{\pm}(k:m)$ for both $m \leq m_P'$ and $m \geq m_P'$, are plotted in Fig. 3. However, due to the duality (\ref{dual}), we have that $\omega_{\pm}^{\rm particle}(k:m) = \omega_{\pm}^{\rm BH}(k:m_P'^2/m)$, and the curves corresponding to particles with $m < m_P'$ are overlaid by the curves corresponding to black holes with masses $m_P'^2/m >m_P'$. Each curve in the first set, labelled by the mass $m$, is overlaid by a curve in the second set, labelled by $m_P'^2/m$, and each quadrant contains {\it both} ``particle" and ``black hole" solutions. The lower-left, lower-right, upper-left and upper-right quadrants corresponds to the four independent regimes, $\omega_{-} \leq \omega_P$, $k \leq k_P$; $\omega_{-} \leq \omega_P$, $k \geq k_P$; $\omega_{+} \geq \omega_P$, $k \leq k_P$ and $\omega_{+} \geq \omega_P$, $k \geq k_P$, respectively. Therefore, only the lower left-hand quadrant is {\it definitely} physical. It describes both quantum particles and quantum black holes with super-Planckian radii. The canonical QM dispersion relation, $\omega = (\hbar/2m)k^2$, corresponds to the asymptotic regime in the bottom left-hand corner of the diagram.

{\it If} the upper half-plane of Fig 3. is regarded as physical, the upper right-hand quadrant gives the dispersion relation for ``sub-Planckian" black holes -- that is, black holes with $\lambda_S < l_P$, $m < m_P$, conjectured to exist in LQG \cite{Carr:2015nqa} -- or, equivalently, the dispersion relation for ``sub-Planckian" particles -- that is, particles with $\lambda_C < l_P$, $m > m_P$. If the latter exist, they may correspond to the point-like matter distributions at the centre of normal (super-Planckian) black holes, thereby preventing the formation of classical singularities. The upper left-hand and lower right-hand quadrants correspond to mixed regimes, where objects are associated with either sub-Planckian time periods and super-Planckian wavelengths, or vice-versa. We note that all four solution branches ``touch" in the critical case, $m=m_P'$, $\omega = \omega_P$, $k=k_P$, which corresponds to a {\it unique} quantum state in the semi-Newtonian picture. The limits $k_{\pm}(m)$ are shown in Fig. 4. (Both Figs. 3-4 are taken from \cite{Lake:2015pma}.)

\begin{figure}[h] \label{Fig.3}
\centering
\includegraphics[width=10cm]{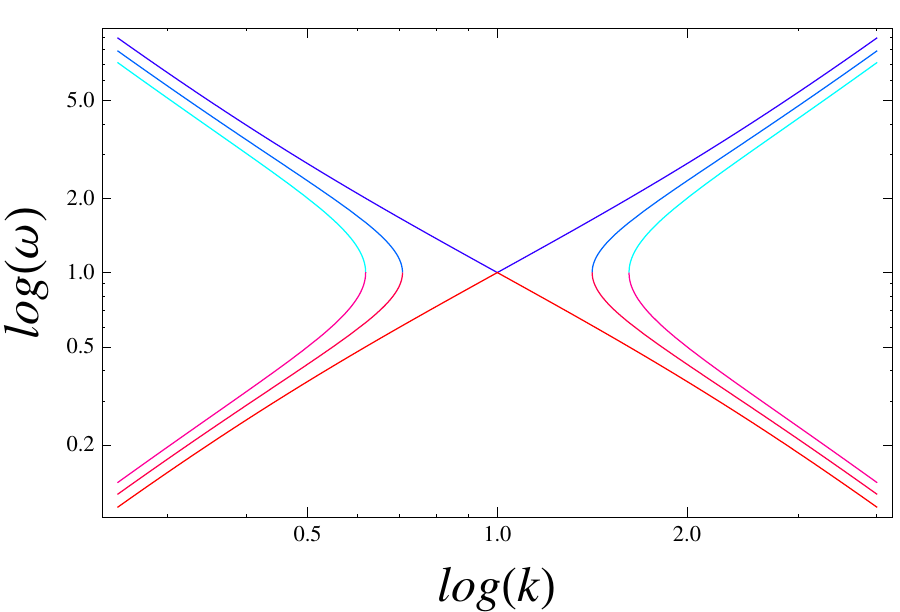}
\caption{$\omega_{+} (k:m)$ (blue) and $\omega_{-} (k:m)$ (red), as a function of $k$, for two mass values in the range $m < m_P'$ (or, equivalently, their duals in the range $m > m_P'$), together with the critical value $m = m_P'$. In this and all subsequent plots, we choose units such that $\omega_P = k_P = 1$, \cite{Lake:2015pma}.}
\end{figure}

\begin{figure}[h] \label{Fig.4}
\centering
\includegraphics[width=10cm]{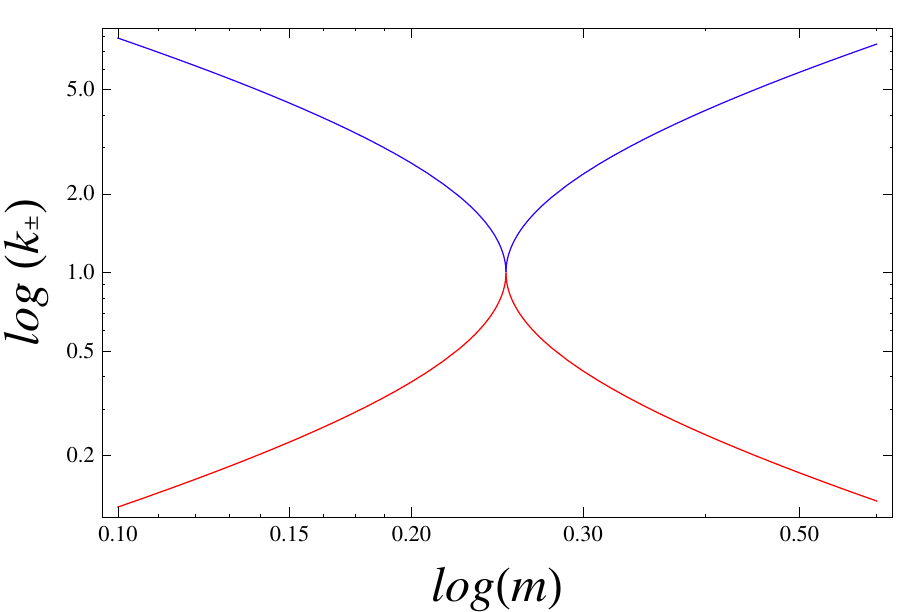}
\caption{Upper/lower limits on the wave number, $k_{+}$ (blue) and $k_{-}$ (red), as functions of $m$, \cite{Lake:2015pma}.}
\end{figure}

\section{Operators and EOM for the modified theory} \label{Sec.5}

In this section, we determine the EOM for the quantum system, implied by the extended de Broglie relations proposed in Sec. \ref{Sec.4}, and the corresponding Hamiltonian and momentum operators of the modified theory.

\subsection{Basic definitions} \label{Sec.5.1}

For $m \leq m_P'$, Eqs. (\ref{Omega2})-(\ref{kappa2}) suggest the following definitions for the Hamiltonian and momentum operators, and the quantum EOM: 
\begin{eqnarray} 
\label{H:m<m_P:w<w_p:k<k_P}
\hat{H}\psi = \frac{\hat{p}^2}{2m}\psi \equiv -\frac{\hbar^2}{2m}[1 - k_P^{-2}(\partial^2 /\partial x^2)]^{-2}\frac{\partial^2 \psi}{\partial x^2} = i\hbar[1 - \omega_P^{-2}(\partial^2 /\partial t^2)]^{-1}\frac{\partial \psi}{\partial t} \, .
\end{eqnarray} 
Equation (\ref{H:m<m_P:w<w_p:k<k_P}) is obtained by substituting the standard definitions of the ``wave number operator" and ``angular frequency operator"
\begin{eqnarray}
\hat{k} = -i(\partial/\partial x) \, , 
\nonumber\\
\hat{\omega} = +i(\partial/\partial t) \, ,
\end{eqnarray} 
which yield $\hat{\omega}\phi = \omega \phi$, $\hat{k}\phi = k\phi$, where $\phi(x,t) = e^{i(kx - \omega t)}$, into the modified de Broglie relations. Though the meaning of the operators inside the square brackets is unclear, we note that they may be defined as Taylor series by analogy with standard algebraic quantities, giving
\begin{eqnarray} \label{Exp1A}
[1 - k_P^{-2}(\partial^2 /\partial x^2)]^{-2} = \sum_{n=0}^{\infty}(n+1)k_P^{-2n}\frac{\partial^{2n}}{\partial x^{2n}} \, , 
\end{eqnarray} 
\begin{eqnarray} \label{Exp1B}
[1 - \omega_P^{-2}(\partial^2 /\partial t^2)]^{-1} = \sum_{n=0}^{\infty}\omega_P^{-2n}\frac{\partial^{2n}}{\partial t^{2n}} \, .
\end{eqnarray} 
Strictly, Eqs. (\ref{Exp1A})-(\ref{Exp1B}) are valid only when applied to wave packets composed of plane-wave modes with $\omega < \omega_P$ and $k < k_P$, since
\begin{eqnarray} \label{Exp1C}
\sum_{n=0}^{\infty}(n+1)k_P^{-2n}\frac{\partial^{2n}\phi}{\partial x^{2n}} &=& \sum_{n=0}^{\infty}(n+1)k_P^{-2n}(ik)^{2n}\phi = [1 + (k/k_P)^2]^{-2}\phi \, , 
\end{eqnarray} 
\begin{eqnarray} \label{Exp1D}
\sum_{n=0}^{\infty}\omega_P^{-2n}\frac{\partial^{2n}\phi}{\partial t^{2n}}  &=& \sum_{n=0}^{\infty}\omega_P^{-2n}(-i\omega)^{2n}\phi = [1 + (\omega/\omega_P)^2]^{-1}\phi \, ,
\end{eqnarray} 
and these series converge for $k/k_P < 1$, $\omega/\omega_P < 1$, respectively. In the $\omega \ll \omega_P$ and $k \ll k_P$ regime, the higher order terms are subdominant and we obtain the standard Schr{\" o}dinger equation. However, for $\omega \approx \omega_P$ and $k \approx k_P$, which \emph{necessarily} applies for any particle with rest mass $m \approx m_P'$, higher order corrections become significant. 

This is an important point. Even in the canonical QM, we are not free to consider a quantum state with $\omega \approx \omega_P$ and $k \approx k_P$ for arbitrary $m$, since these values necessarily imply $E \approx 2\pi m_Pc^2$, $m \approx \pi m_P$, according to the standard dispersion relation (\ref{non_rel_disp}). Similarly, in the extended theory, the limit $m \rightarrow m_P'$ implies $\omega \rightarrow \omega_P$, $k \rightarrow k_P$ and vice-versa. The point $(\omega_P,k_P)$ is inaccessible for particles with $m \neq m_P'$, due to the existence of the limiting values $k_{\pm}$(m). As stated in Sec. \ref{Sec.4.2}, systems with mass $m_P'$, be they interpreted as particles or black holes, are \emph{unique} in this prescription. Alternatively, they may be interpreted as black hole relics \cite{Barrow:1992hq} formed at the end of Hawking evaporation \cite{Hawking:1974rv,Bardeen:1973gs}. If they exist, such relics represent unique quantum and gravitational states, in which the Schwarzschild and Compton radii coincide. 

For $m \geq m_P'$, the quantum EOM may be written as
\begin{eqnarray}  \label{H:m>m_P:w<w_p:k<k_P}
\hat{H}'\psi = \frac{\hat{p}'^2}{2m}\psi \equiv -\frac{mc^2}{k_P^2}[1 - k_P^{-2} (\partial^2 /\partial x^2)]^{-2}\frac{\partial^2 \psi}{\partial x^2} = \frac{i\hbar}{8} [1 -  \omega_P^{-2}(\partial^2 /\partial t^2)]^{-1}\frac{\partial \psi}{\partial t} \, .
\end{eqnarray} 
where $\hat{H}'$ and $\hat{p}'$ denote redefined Hamiltonian and momentum operators, that do {\it not} reduce to canonical form for $\omega \ll \omega_P$, $k \ll k_P$. Instead, in this limit, Eq. (\ref{H:m>m_P:w<w_p:k<k_P}) reduces to 
\begin{eqnarray}  \label{}
\hat{H}'\psi = \frac{\hat{p}'^2}{2m}\psi \equiv -\frac{mc^2}{k_P^2}\frac{\partial^2 \psi}{\partial x^2} \approx \frac{i\hbar}{8}\frac{\partial \psi}{\partial t} \, .
\end{eqnarray} 
This is equivalent to (\ref{Schrod_1D}), with $\beta = 1/4$, as expected.

\subsection{Unitary evolution of self-gravitating states} \label{Sec.5.2}

As both (\ref{H:m<m_P:w<w_p:k<k_P}) and (\ref{H:m>m_P:w<w_p:k<k_P}) contain higher order time derivatives, the issue of unitarity arises. We now demonstrate, explicitly, that the time evolution implied by the extended de Broglie relations (\ref{Omega2})-(\ref{kappa2}) is unitary, at least in the lower-half plane of the $(M,R)$ diagram. This corresponds to the $\omega_{-}(k:m)$ branch of the quantum dispersion relation and to the physically interesting cases of both self-gravitating particles, and black holes, in the semi-Newtonian approximation. 

In canonical QM, the Schr{\" o}dinger equation with \emph{general} time-independent Hamiltonian $\hat{H}(x)$ is 
\begin{eqnarray} \label{Schrod_QM}
i\hbar\frac{\partial \psi}{\partial t} = \hat{H}(x)\psi \, .
\end{eqnarray} 
The time evolution of a single momentum eigenstate, $\phi(x,0) = e^{ikx}$, is therefore given by the unitary operator
\begin{eqnarray} \label{U_k_QM}
\hat{U}_k(t) = e^{-i\omega(k)t}\mathbb{I} = e^{-iE_kt/\hbar}\mathbb{I} \, ,
\end{eqnarray}
where $\hat{H}(x)e^{ikx} = E_ke^{ikx}$ and $E_k \equiv E_{\omega(k)} = \hbar\omega(k)$. This ensures that the first de Broglie relation, $E = \hbar\omega$, holds for each individual eigenfunction $\phi(x,t) = e^{i(kx - \omega(k)t)}$, corresponding to the momentum eigenvalue $k$, but the form of $\omega(k)$ cannot be determined without specifying the Hamiltonian. For general wave packets,
\begin{eqnarray} \label{wavepacket_QM}
\psi(x,t) = \int_{-\infty}^{\infty}a(k)e^{i(kx-\omega(k)t)}dk \, ,
\end{eqnarray}
Eq.~(\ref{Schrod_QM}) yields 
\begin{eqnarray} \label{wavepacket_subs_QM}
\hbar\int_{-\infty}^{\infty}a(k)\omega(k)e^{i(kx-\omega(k)t)}dk = \int_{-\infty}^{\infty}a(k)\hat{H}(x)e^{i(kx-\omega(k)t)}dk \, ,
\end{eqnarray}
and it follows from the definition of a function of an operator \cite{Ish95} that the time evolution of a general wave packet, $\psi (x,0)  = \int_{-\infty}^{\infty}a(k)e^{ikx}dk$, is given by 
\begin{eqnarray} \label{U_QM}
\hat{U}(x,t) = e^{-i\hat{H}(x)t/\hbar}\mathbb{I} \, .
\end{eqnarray}
The canonical Hamiltonian for a {\it free} particle is $\hat{H}(x) = -\hbar^2/(2m)(\partial^2/\partial x^2)$. This ensures that the second de Broglie relation, $p = \hbar k$, holds in conjunction with the non-relativistic energy-momentum relation (\ref{non_rel_E-p}), giving the canonical dispersion relation (\ref{non_rel_disp}). 

We now consider the time evolution operator implied by Eq.~(\ref{H:m<m_P:w<w_p:k<k_P}), which we first rewrite as 
\begin{eqnarray} \label{H:m<m_P:w<w_p:k<k_P-Appendix}
i\hbar [ 1 - \omega_P^{-2}(\partial^2 /\partial t^2) ]^{-1}\frac{\partial \psi}{\partial t} = \hat{H}(x)\psi \, ,
\end{eqnarray} 
where
\begin{eqnarray} \label{H:m<m_P:w<w_p:k<k_P_H(x)}
\hat{H}(x) = -\frac{\hbar^2}{2m}[1 - k_P^{-2}(\partial^2 /\partial x^2)]^{-2}\frac{\partial^2 }{\partial x^2} \, .
\end{eqnarray} 
This implies the following time evolution operator, for the single momentum eigenstate $\phi(x,0) = e^{ikx}$, corresponding to the $\omega_{-}(k:m)$ solution branch:
\begin{eqnarray} \label{U_k_1}
\hat{U}_k^{-}(t) = \exp\left[{-i\omega_{-}(k)t}\right]\mathbb{I} = \exp\left[-\frac{i\hbar \omega_P^2}{2E_k}\left(1 - \sqrt{1-\frac{4E_k^2}{\hbar^2 \omega_P^2}}\right)t\right]\mathbb{I} \, ,
\end{eqnarray}
where $E_k \equiv E_{\omega(k)} = \hbar\omega_P^2[\omega(k) + \omega_P^2/\omega(k)]^{-1}$. Note that Eq.~(\ref{U_k_1}) holds for \emph{any} Hamiltonian yielding $\hat{H}(x)e^{ikx} = E_{\omega(k)}e^{ikx}$, not just the one given in (\ref{H:m<m_P:w<w_p:k<k_P_H(x)}). If, in addition, we invoke the definition of the Hamiltonian given by Eq.~(\ref{H:m<m_P:w<w_p:k<k_P_H(x)}), we obtain the modified dispersion relation (\ref{disp_rel_m<m_P1}). 

However, in the modified theory, a subtlety arises when dealing with wave packets, since the integral over $dk$ in Eq. (\ref{wavepacket_QM}) is not well defined everywhere in the range $k \in (-\infty,\infty)$, due to the existence of the limits $k_{\pm}(m)$. We therefore replace (\ref{wavepacket_QM}) with 
\begin{eqnarray} \label{wavepacket_ext}
\psi(x,t) = \int_\Sigma a(k)e^{i(kx-\omega(k)t)}dk \, ,
\end{eqnarray}
where $\Sigma \subset \mathbb{R}$ is the appropriate range for the quadrant considered. For example, for wave packets governed by Eq.~(\ref{H:m<m_P:w<w_p:k<k_P-Appendix}), the integration must be performed over $\Sigma = [-k_{-}(m),k_{-}(m)]$, where $k_{-}(m)$ is given by Eq. (\ref{disp_rel_m<m_P4}). Substituting (\ref{wavepacket_ext}) into (\ref{H:m<m_P:w<w_p:k<k_P-Appendix}) gives
\begin{eqnarray} \label{wavepacket_subs_1}
\hbar\omega_P^2\int_{-k_-(m)}^{k_-(m)} a(k)[\omega(k) + \omega_P^2/\omega(k)]^{-1}e^{i(kx-\omega(k)t)}dk = \int_{-k_-(m)}^{k_-(m)}
a(k)\hat{H}(x)e^{i(kx-\omega(k)t)}dk \, .
\end{eqnarray}
Again using the definition of a function of an operator \cite{Ish95}, the time evolution of a general wave packet, $\psi(x,0) = \int_{\Sigma}a(k)e^{ikx}dk$, is 
\begin{eqnarray} \label{U_1}
\hat{U}^{-}_{m \leq m_P'}(x,t) = \exp\left[-\frac{i \hbar \omega_P^2}{2\hat{H}_{-}(x)}\left(1 - \sqrt{1-\frac{4\hat{H}_{-}^2(x)}{\hbar^2 \omega_P^2}}\right)t\right]\mathbb{I} \, .
\end{eqnarray}
For clarity, we have relabelled the Hamiltonian (\ref{H:m<m_P:w<w_p:k<k_P_H(x)}), which is valid for the negative solution branch, $\omega_{-}(k:m)$, for $m \leq m_P'$, as $\hat{H}_{-}(x)$. Thus, the evolution is \emph{non-canonical but unitary}. 

The operator (\ref{U_1}) gives the time evolution of a fundamental, self-gravitating, quantum mechanical particle. Using Eqs.~(\ref{H:m<m_P:w<w_p:k<k_P_H(x)}) and (\ref{Exp1A}), we see that, if $E_k \ll (1/2)\hbar\omega_P$, which is equivalent to taking the limit $m \ll m_P'$, we have $\omega_{-}(k:m) \approx E_k/\hbar \approx \hbar k^2/(2m)$. Thus, Eq.~(\ref{U_k_1}) reduces to the standard time evolution for a single eigenstate (\ref{U_k_QM}). When this conditions holds for all $k \in [-k_{-}(m),k_{-}(m)]$, Eq.~(\ref{U_1}) gives the standard time evolution for a wave packet (\ref{U_QM}). In this case, the wave packet expansion (\ref{wavepacket_ext}) also reduces to the standard form (\ref{wavepacket_QM}), since $|k_{-}(m)| \rightarrow \infty$ for $m/m_P' \rightarrow 0$. In other words, for $m \ll m_P'$, corrections due to gravitational effects become negligible and the extended de Broglie theory obeys the time evolution and expansion theorems of canonical QM. For $m \geq m_P'$, an analogous argument implies that the time-evolution operator is
\begin{eqnarray} \label{U_2}
\hat{U}^{-}_{m \geq m_P'}(x,t) = \exp\left[-\frac{i \hbar \omega_P^2}{2\hat{H}_{-}'(x)}\left(1 - \sqrt{1-\frac{4\hat{H}_{-}'^2(x)}{\hbar^2 \omega_P^2}}\right)t\right]\mathbb{I} \, ,
\end{eqnarray}
where $\hat{H}_{-}'(x)$ denotes the (redefined) Hamiltonian which is valid for the negative solution branch, $\omega_{-}(k:m)$, for $m \geq m_P'$. This is defined in Eq. (\ref{H:m>m_P:w<w_p:k<k_P}) and is simply the dual of (\ref{H:m<m_P:w<w_p:k<k_P_H(x)}) under the transformation (\ref{dual}). 

The following symmetry exists between the time evolution operators particles (\ref{U_1}) and black holes (\ref{U_2}):
\begin{eqnarray} \label{symmetries-1}
\hat{U}^{-}_{m \leq m_P'}(x,t) \leftrightarrow \hat{U}^{-}_{m \geq m_P'}(x,t) \iff \hat{H}_{-}(x) \leftrightarrow \hat{H}_{-}'(x) = (m_P'/m)^2\hat{H}_{-}(x) \, . 
\end{eqnarray}
In the limit 
\begin{eqnarray} \label{limit-1}
m \rightarrow m_P' \iff \omega \rightarrow \omega_P, \ k \rightarrow k_P \ \ \forall \omega, k \, ,
\end{eqnarray}
we have $k_{+}(m_P') = k_{-}(m_P') = k_P$, so that
\begin{eqnarray} \label{limit-2}
\psi(x,0) \rightarrow \phi(x,0) = e^{ik_Px} \, .
\end{eqnarray}
This is the \emph{unique} (initial) momentum eigenstate for a particle with mass $m = m_P'$ in the extended de Broglie theory. In this limit, the Hamiltonians and time evolution operators for the $m \leq m_P'$ and $m \geq m_P'$ sectors converge, yielding
\begin{eqnarray} \label{limit-3}
\hat{H}_{-}(x) = \hat{H}_{-}'(x) = 2m_P'c^2 \, 
\end{eqnarray}
and
\begin{eqnarray} \label{limit-4}
\hat{U}^{-}_{m \leq m_P'}(x,t) = \hat{U}^{-}_{m \geq m_P'}(x,t) = e^{-i\omega_P t} \, .
\end{eqnarray}

\section{Implications for the Hawking temperature} \label{Sec.6}

The Hawking temperature \cite{Hawking:1974rv,Bardeen:1973gs} for black holes may also be obtained heuristically in the semi-Newtonian approximation of canonical QM. Assuming an upper limit for $\Delta_{\psi}x$, associated with the Schwarzschild radius, we obtain a corresponding lower limit on $\Delta_{\psi}p$ via the HUP (\ref{HUP}), giving
\begin{eqnarray}\label{}
(\Delta_{\psi}x)_{\rm max} \approx \lambda_S \, , \quad \quad (\Delta_{\psi}p)_{\rm min} \approx \frac{\hbar}{\lambda_S} \approx \frac{m_P^2c}{m} \, .
\end{eqnarray}
$(\Delta_{\psi}p)_{\rm min}$ may be identified with the momentum uncertainty of a particle emitted during Hawking radiation \cite{Ca:2014,cmp}, and the associated temperature $T_H$ is defined via
\begin{eqnarray}\label{}
(\Delta_{\psi}p)_{\rm min} = \chi m_Pc \frac{T_H}{T_P}\, ,
\end{eqnarray}
where $T_P = m_Pc^2/k_B$ is the Planck temperature and $\chi$ is a constant. Agreement with the Hawking temperature requires $\chi = 2 \pi$, so that
\begin{eqnarray}\label{T_H}
T_H = \frac{\hbar c^3}{8\pi Gm k_B} = 
\frac{1}{8\pi}\frac{m_P T_P}{ m} \, .
\end{eqnarray}
In this scenario, we note that, since $T_H$ is associated with a lower limit on the momentum uncertainty, it represents the {\it minimum} temperature that can be associated with the mass $m$ of the black hole. In other words, were an observer to localise $m$ on some scale $\Delta_{\psi}x < \lambda_S$, for example, by crossing the event horizon, we would expect the temperature associated with $m$, in their frame, to be greater than $T_H$.

As shown in Sec. \ref{Sec.2.1}, we may also obtain the standard expression for the Compton wavelength (up to numerical factors of order unity) by associating this with the lower limit on the size of the wave packet $(\Delta_{\psi}x)_{\rm min}$. The corresponding upper limit on the momentum uncertainty $(\Delta_{\psi}p)_{\rm max}$ is associated with the particle's rest mass (\ref{Compton-HUP}). This defines the Compton line on the $(M,R)$ diagram, and the numerical factor may be chosen to ensure that the intersect with the Schwarzschild line occurs at $m = m_P'$ (\ref{Compton_redef}). This allows us to associate the two halves of the diagram with the extended de Broglie relations for particles and black holes, respectively, given in Eqs. (\ref{Omega2})-(\ref{kappa2}).

We then define the ``Compton temperature" as
\begin{eqnarray}\label{T_C}
T_C = \frac{1}{\chi \pi^2}\frac{m}{m_P}T_P = \frac{1}{2 \pi^3}\frac{mc^2}{k_B} \, ,
\end{eqnarray}
where setting $\chi = 2 \pi$ ensures that $T_C = T_H = (2\pi)^{-1}T_P$ at $m = m_P'$. This may also be regarded as the natural temperature associated with the rest mass of the particle. By our previous logic, since $T_C \propto (\Delta_{\psi}p)_{\rm max} \approx mc$ is associated with the upper limit on the momentum uncertainty, it represents the {\it maximum} temperature that can be associated with the particle mass $m$. This matches our intuition, since particles with real temperatures $T \gtrsim T_C$, corresponding to wave packets with $\Delta_{\psi}x \lesssim \lambda_C$, have sufficient kinetic energy to pair produce. Additional increases in the total energy of the system then result in pair production rather than increased temperature. Conversely, wave packets with $\Delta_{\psi}x \gtrsim \lambda_C$ are naturally associated with temperatures $T \lesssim T_C$, which correspond to kinetic energy uncertainties $\Delta E \lesssim mc^2$. 

$T_C$ and $T_H$ are plotted as red and blue lines, respectively, in Fig. 5, which is also taken from \cite{Lake:2015pma}. The green curve $T_{C/H}'$ represents a smooth interpolation between the two, which is associated with a unified Compton-Schwarzschild line, $\lambda_{C/S}'$, via $T_{C/H}' \propto (\Delta_{\psi}p)_{\rm max/min} \propto \hbar/\lambda_{C/S}'$, by analogy with the semi-Newtonian ``derivations" of $T_C$ and $T_H$ in the asymptotic regions, given above. Just as the physical nature of the radius changes at the critical mass, $m = m_P'$, as illustrated explicitly in Fig. 2, so does the physical nature of the temperature. The left-hand and right-hand halves of the green curve in Fig. 5 may be associated with upper and lower limits on $T$, respectively, so that the $(T,m)$ diagram analogue of Fig. 2 would appear inverted, with shaded and unshaded regions interchanged.   

\begin{figure}[h] \label{Fig.5}
\centering
\includegraphics[width=12cm]{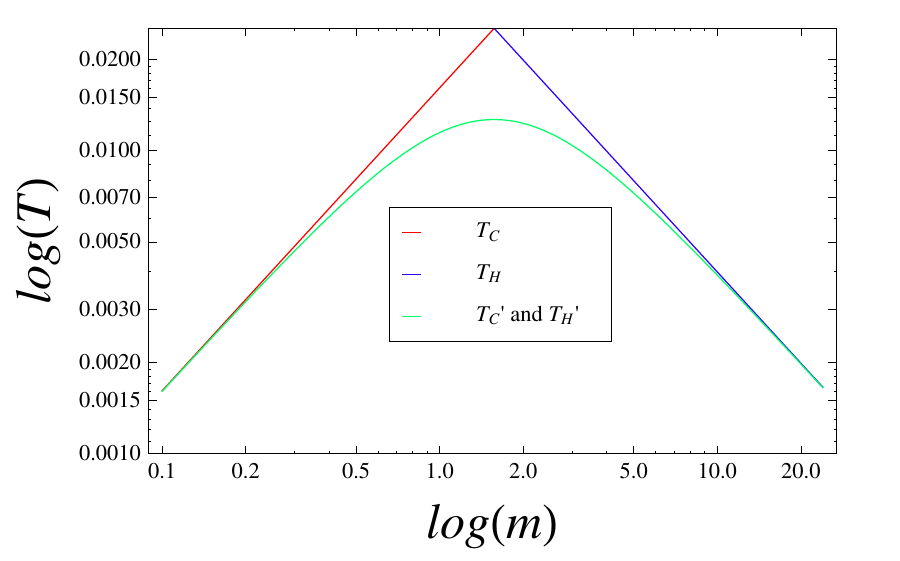}
\caption{The Compton temperature $T_C$, as defined in the semi-Newtonian approximation of canonical QM (red line), and Hawking temperature $T_H$, derived in semi-classical gravity (blue line). $T_{C/H}'$ represents a unified temperature curve, suggested by the theory of extended de Broglie relations (green curve) \cite{Lake:2015pma}.}
\end{figure}

\section{Unification of black holes and fundamental particles from GUP phenomenology and its relation to extended de Broglie relations} \label{Sec.7}

As discussed in Sec. \ref{Sec.2.1}, substituting the commutator $[\hat{x},\hat{p}] = i\hbar$ into the general uncertainty principle (\ref{SUP}) yields the famous Heisenberg uncertainty principle (HUP) $\Delta_{\psi}x \Delta_{\psi}p \geq \hbar/2$, which arises as a fundamental consequence of the Hilbert space structure of canonical QM. Although $\Delta_{\psi}x$ and $\Delta_{\psi}p_x$ do \emph{not} refer to any unavoidable ``noise" , ``error" or  ``disturbance" introduced to the system by the act of measurement, it is well known that one can understand this result {\it heuristically} as reflecting the momentum transferred to the particle by a probing photon \cite{Rae00}. Hence, in order to distinguish between quantities representing genuine noise, induced by the measurement process, and the standard deviations of repeated {\it perfect}, projective, Von Neumann-type measurements, which do not disturb the state $\ket{\psi}$ prior to wave function collapse \cite{Ish95}, we use the notation $\Delta O$ for the former and $\Delta_{\psi} O$ for the latter. (Strictly speaking, any disturbance to the state of the system caused by an act of measurement may also be $|\psi\rangle$-dependent, but we adopt Heisenberg's original notation, in which the state-dependent nature of the disturbance is not explicit.)
In this notation, the formulation of the standard uncertainty principle based on the ``Heisenberg microscope" thought experiment may be written as
\begin{eqnarray} \label{HUP_xp}
\Delta x\Delta p \gtrsim \hbar \, .
\end{eqnarray}
We note that such a statement must be viewed as a postulate with no rigorous foundation in the underlying mathematical structure of quantum theory. Indeed, as a postulate, it has recently shown to be manifestly false, both theoretically \cite{Oz03A,Oz03B} and experimentally \cite{Roetal12,Eretal12,Suetal13,Baetal13}. However, the heuristic ``derivation" of Eq. (\ref{HUP_xp}) may be found in many older texts, alongside the more rigorous derivation of Eq. (\ref{SUP}) from basic mathematical principles. Unfortunately, it is not always made clear that the quantities involved in each expression are different, as clarified by the recent pioneering work of Ozawa \cite{Oz03A,Oz03B}. An excellent discussion regarding the various possible meanings and (often confused) interpretations of symbols like ``$\Delta x$" is given in \cite{Scheibe}.

The clarifications given above are of key importance when discussing quantum phenomenology based on the Generalised Uncertainty Principle (GUP). This term refers, generically, to {\it any} extension of the the standard HUP (\ref{HUP_xp}), whether based on fundamental modifications of the mathematical formalism of canonical QM, or simply on heuristic results. It must not be confused with the term ``general uncertainty principle" used to describe Eq. (\ref{SUP}), though the terminology adopted in different parts of the literature is confusing \cite{Ish95}. 

Many GUP proposals are motivated by attempts to incorporate gravitational effects into existing interpretations of quantum uncertainty. Many also give rise to a minimum length, interpreted as a quantum gravitational effect, and are known as minimum length uncertainty relations (MLURs). (For recent reviews of GUP phenomenology, see \cite{Tawfik:2014zca,Tawfik:2015rva}; for reviews of minimum length scenarios in quantum gravity, see \cite{Hossenfelder:2012jw,Garay:1994en}.) One of the earliest and most interesting examples of a quantum gravity inspired GUP was proposed by Adler and Santiago \cite{Adler:1999bu},
\begin{eqnarray}\label{AdlerGUP}
\Delta x \Delta p \gtrsim \hbar + \alpha (\Delta p)^2 \, ,
\end{eqnarray}
where $\alpha$ is a constant of order $l_P^2/\hbar \sim G/c^3$. This is based on an extension of the Heisenberg microscope thought experiment, incorporating the effect of the gravitational interaction between the particle and the photon probe, which gives rise to the additional term on the right-hand side. A GUP of this form was originally proposed in the context of string theory \cite{Amati:1988tn}, as it implies the existence of a minimum length, $(\Delta x)_{\rm min} \approx l_P + \alpha \hbar/l_P \approx l_P$, which may be identified with the fundamental string scale. 

From (\ref{AdlerGUP}), it is clear that using the standard identification for the momentum uncertainty $\Delta p \rightarrow mc$ implies $\Delta x \rightarrow \lambda_C'$, where $\lambda_C'$ represents a modified Compton wavelength of the form 
\begin{eqnarray}\label{GUP_Compton}
\lambda_C' \approx \frac{\hbar}{mc} + \frac{Gm}{c^2} \approx \lambda_C + \lambda_S \, .
\end{eqnarray}
Thus, for $m \lesssim m_P$, the first (Compton) term dominates, whereas, for $m \gtrsim m_P$, the second (Schwarzschild) term gives the leading order contribution to the total uncertainty. In the this way, a GUP of the form (\ref{AdlerGUP}) may enable a unified description of fundamental particles and black holes based on a modified quantum theory. A GUP-based unification of this this kind, suggested in \cite{Ca:2014}, is termed the black hole uncertainty principle (BHUP) correspondence. However, several theoretical and conceptual issues arise in the context of such a proposal, which we now address.

First, if we wish to give rigorous mathematical definitions of the terms $\Delta x$ and $\Delta p$ appearing in Eq. (\ref{AdlerGUP}) (i.e., to ``promote" them to quantities analogous to $\Delta_{\psi} x$ and $\Delta_{\psi} p$, which follow directly from the mathematical formalism of the underlying quantum theory (\ref{SUP})), it is clear that we must modify either the Hilbert space structure of canonical QM and/or the commutator $[\hat{x},\hat{p}] = i\hbar$. The former is highly non-trivial, while the latter is equivalent to leaving the state space structure unchanged but modifying the dispersion relations and the EOM for the state vector. This, in turn, may be achieved in several different ways: (i) by maintaining the standard de Broglie relations but modifying the classical energy-momentum relation, (ii) by maintaining the classical energy-momentum relation but modifying the de Broglie relations, and (iii) by modifying both. One approach, based on (ii), gives rise to the modified commutator \cite{Hossenfelder:2003jz}
\begin{eqnarray}\label{Hossenfelder-1}
[\hat{x},\hat{p}] \approx i\hbar \left(1 + \alpha \hat{p}^2\right) \, ,
\end{eqnarray}
which corresponds to the individual operators
\begin{eqnarray}\label{Hossenfelder-2}
\hat{x} \approx i\hbar \left(1 + \frac{\alpha}{\hbar} \hat{p}^2\right)\frac{\partial}{\partial p} \, , \quad \hat{p} \approx i\hbar \left(1 - \frac{\alpha}{3} \frac{\partial^2}{\partial x^2}\right)\frac{\partial}{\partial x} \, . 
\end{eqnarray}
Substituting (\ref{Hossenfelder-1}) into (\ref{SUP}) and setting $\langle p \rangle \approx 0$, so that $\Delta_{\psi}p \approx \sqrt{\langle p^2 \rangle}$, yields a GUP of the form 
\begin{eqnarray}\label{Hossenfelder-3}
\Delta_{\psi} x \Delta_{\psi} p \gtrsim \frac{\hbar}{2}\left[1 + \frac{\alpha}{\hbar} (\Delta_{\psi} p)^2\right] \, ,
\end{eqnarray}
which is obviously equivalently to (\ref{AdlerGUP}), up to numerical factors of order unity, but which may be derived from first principles. Crucially, the approximation $\langle p \rangle \approx 0$ is valid for spherically symmetric systems, in which it is also reasonable to make the identification $\Delta_{\psi} p \approx mc$. (In fact, it may be argued that identifying the momentum uncertainty with the rest mass is valid {\it only} in the rest frame of the system, under conditions of spherical symmetry.) Hence, the modified commutator (\ref{Hossenfelder-1}) is {\it potentially} capable of yielding a unified Compton-Schwarzschild line of the form (\ref{GUP_Compton}) under the identification $\Delta_{\psi} p \approx mc$, but, for reasons we now address, such an identification is not possible for arbitrary $m$.

In \cite{Hossenfelder:2003jz}, the modified dispersion relations giving rise to Eqs. (\ref{Hossenfelder-1})-(\ref{Hossenfelder-3}) are
\begin{eqnarray}\label{Hossenfelder-4}
k(p) = \frac{1}{l_P}\tanh\left(\frac{p}{m_Pc}\right) \approx \frac{p}{\hbar} - \frac{1}{3l_P}\left(\frac{p}{m_Pc}\right)^3 + \dots \, ,  
\nonumber\\
\omega(E) = \frac{c}{l_P}\tanh\left(\frac{E}{m_Pc^2}\right) \approx \frac{E}{\hbar} - \frac{1}{3l_P}\left(\frac{E}{m_Pc^2}\right)^3 + \dots \, , 
\end{eqnarray}
where the approximations on the right-hand sides of Eqs. (\ref{Hossenfelder-4}) are valid for $p \ll m_Pc$ and $E \ll m_Pc^2$, respectively. If these hold for every de Broglie mode in a single spherically symmetric wave packet, then $\Delta_{\psi} p \approx mc \ll m_Pc$. Hence, this identification is incapable of providing a unified description of particles and black holes, as it does not allow for the extension of the modified Compton wavelength $\lambda_C'$ (\ref{GUP_Compton}) into the right-hand side of the $(M,R)$ diagram. 

However, other considerations suggest an alternative identification between $\Delta_{\psi} p$ and the rest mass of a spherically symmetric system for $m \gtrsim m_P$. Consider a black hole decaying via Hawking radiation: The average magnitude of the momentum of a particle emitted from a black hole of mass $m_{\rm BH}$ is $\sqrt{\langle p_{\rm particle}^2 \rangle} \approx (m_P^2/m_{\rm \rm BH})c \equiv m_{\rm particle}c$, where $m_{\rm particle}$ is the effective particle mass (even if the particle is a photon). Therefore, $(\Delta_{\psi} p)_{\rm particle} \approx (m_P^2/m_{\rm BH})c$ is the total momentum uncertainty; this may also be seen by associating the positional uncertainty of the emitted particle with the black hole radius, $(\Delta_{\psi} x)_{\rm particle} \approx Gm_{\rm BH}/c^2$, in the standard HUP, or in the leading order term of the GUP (\ref{AdlerGUP}) for particles with $m_{\rm particle} \ll m_P$. Since momentum conservation implies that the recoil of the black hole, due to a {\it single} emission, must be equal in magnitude (but opposite in sign) to the momentum of the outgoing particle, it follows that $(\Delta_{\psi} p)_{\rm particle} = (\Delta_{\psi} p)_{\rm BH} \approx (m_P^2/m_{\rm BH})c$. Thus, for general spherically symmetric systems of mass $m$ -- subject to the HUP or a GUP of the form (\ref{AdlerGUP}) -- we may associate
\begin{equation} 
\label{mass-assoc}
\Delta_{\psi} p \approx \left \lbrace
\begin{array}{rl}
mc \, ,  & \ (m \lesssim m_P) \\
(m_P^2/m)c \, ,  & \ (m \gtrsim m_P) \, . 
\end{array}\right.
\end{equation}
An interesting feature of the GUP proposed by Adler and Santiago (\ref{AdlerGUP}) is that it is {\it invariant} under the transformation $\Delta_{\psi} p \rightarrow m_Pc^2/\Delta_{\psi} p$, so that the change in how $\Delta_{\psi} p$ is associated with the rest mass of the system, occurring at $m \approx m_P$, still enables the construction of a unified Compton-Schwarzschild line of the form (\ref{GUP_Compton}). 

We also note that the modified commutator and dispersion relations proposed in \cite{Hossenfelder:2003jz} do not alter the fundamental Newtonian nature of space and time (subject to the cosmic speed limit $v_{\rm max} = c$), since they leave the fundamental commutator between the $\hat{x}$ and $\hat{k}$ operators unchanged, 
\begin{eqnarray}\label{Hossenfelder-5}
[\hat{x}^i,\hat{k}_j] = i\delta^{i}{}_{j} \, ,
\end{eqnarray}
where 
\begin{eqnarray}\label{Hossenfelder-6}
\hat{x}^i = i\frac{\partial}{\partial k_i} \, , \quad \hat{k}_i = -i\frac{\partial}{\partial x^i} \, .
\end{eqnarray}
What has changed is, instead, is the the intrinsic relation between $p$ and $k$, i.e. the functional form of $p(k)$. Thus, in this case, the generators of infitesimal translations in the Gallilean symmetry group may still be identified with the $\hat{k}_i$ operators; translation invariance gives rise to conserved Noether charges for {\it classical} systems, the components of the classical momentum $p_i$, {\it and} to a set of nontrivial commutators (\ref{Hossenfelder-5}) for quantum systems. However, the hermitian operators $\hat{p}_i$ are no longer directly proportional to the de Broglie wave number operators $\hat{k}_i$, due to the more complex form of the modified de Broglie relations (\ref{Hossenfelder-5}). As with the extended de Broglie relations presented in Secs. \ref{Sec.4}-\ref{Sec.6}, these remain the essence of the non-trivial modifications to the quantum sector. (Similar arguments apply regarding time translational symmetry, the frequency operator $\hat{\omega} = i(\partial/\partial t)$, and the classical energy $E$, though subject to the caveat that time $t$ is parameter, {\it not} an operator, in both canonical QM and the modified theories presented here.)  

Hence, considering the GUP, together with momentum conservation arguments that give rise to the approximate identifications (\ref{mass-assoc}) in the two mass regimes, is phenomenologically equivalent to using the HUP together with the associations
\begin{eqnarray}\label{mass-assoc*}
\Delta_{\psi} x \approx \frac{l_P}{m_P}\left(m + m_P^2/m\right) \, , \quad \Delta_{\psi} p \approx \frac{m_P^2c}{m + m_P^2/m} \, , 
\end{eqnarray}
which are assumed to be valid for {\it all} $m$. This makes sense, intuitively, since it relates the positional and momentum uncertainty of quantum-gravitational system to the ADM mass \cite{Carr:2015nqa,Arnowitt:1959ah}. We also note that Eq. (\ref{mass-assoc*}) gives rise to a unified Compton-Schwarzschild line that is (at least) {\it qualitatively} similar to the effective mass-dependence of $\lambda_{C/S}'$, introduced via the extended de Broglie theory given in Secs. \ref{Sec.4}-\ref{Sec.6}. In the latter, however, the rest mass is directly associated with the minimum uncertainty of a {\it modified} momentum operator $(\Delta_{\psi}p')_{\rm min} \approx mc$ (\ref{x_max_m>>m_P}). The arguments presented here suggest that this {\it cannot} be identified with the physical recoil of the black hole during the Hawking emission of a fundamental particle. 

Another crucial difference between the two approaches is that the modified de Broglie relations proposed in \cite{Hossenfelder:2003jz}, Eqs. (\ref{Hossenfelder-4}), predict an {\it almost} unique quantum state for {\it all} black holes, since $k(p) \rightarrow 1/l_P$ for $p \gg m_Pc$ and $\omega(E) \rightarrow c/l_P$ for $E \gg m_Pc^2$. This is a direct result of the fundamental asymmetry between the form of the relations (\ref{Hossenfelder-4}) for $m \lesssim m_P$ and $m \gtrsim m_P$. By contrast, the extended de Broglie theory presented in Secs. \ref{Sec.4}-\ref{Sec.6} explicitly incorporates the symmetry between fundamental particles and black holes present in the $(M,R)$ diagram, via the extensions of the relations themselves. In this scenario, the unique quantum state $\phi(x:t) = \exp[i(k_P x - \omega_P t)]$ corresponds to the unique physical state for which Schwarzschild radius is equal to the Compton wavelength. 

A related point concerns the fact that {\it all} forms of modified quantum mechanics proposed thus far in the literature -- be they based directly on GUP phenomenology or on modified de Broglie relations -- give rise to uncertainty relations of the form $\Delta_{\psi} x \gtrsim \dots $, even when these are extended to systems with super-Planckian rest mass. This is the mathematical expression of the assumption that the radius of a black hole is, in some sense, physically equivalent to the radius of fundamental particle, i.e. that it may be considered as a minimum, rather than as a maximum limit on rest mass localisation. Were $\psi$ to represent (in some limit) the wave function associated with the mass in the interior of a black hole, it may be argued that a modified quantum theory giving rise to a positional uncertainty of the form $\Delta_{\psi} x \lesssim \dots $ is required. It would be interesting to consider how such a theory could be reconciled with the ``horizon wave function" model, proposed by Casadio {\it et al} (see, for example \cite{Casadio:2015qaq,Casadio:2013tza,Casadio:2013aua,Casadio:2014vja}), as even an effective wave function model of a black hole may require {\it both} components. 

\section{Conclusions} \label{Sec.8}

We have shown that, by taking the semi-Newtonian approximation, in which space and time remain absolute but a cosmic speed limit  $v_{\rm max} = c$ is imposed ``by hand", we can recover essential phenomenological features of both relativistic quantum theory and relativistic gravity from their non-relativistic counterparts. Applying this approximation to the basic mathematical structure of canonical QM yields the formula for the Compton wavelength, via the uncertainty principle, while applying it to Newtonian gravity yields the Schwarzschild radius, via the formula for the escape velocity. On a plot of mass versus radius, these lines meet at the Planck point, suggesting that the final theory of quantum gravity should yield a unified description of fundamental particles and black holes. In the absence of such a theory, we propose a na{\" i}ve unification, based on recognition of the fact that both sets of objects may be adequately described in the semi-Newtonian regime.   

Since the canonical QM dispersion relations break down for Planck mass objects, our new theory is based on modified de Broglie relations, which, when substituted into the classical energy-momentum relation, yield quantum dispersion relations that hold for systems of arbitrary mass. These, in turn, give rise to modified EOM, Hamiltonian, momentum and time-evolution operators, in both the particle and black hole sectors. In each sector, the time-evolution is non-canonical, but unitary, and the particle evolution reduces to canonical form in the asymptotic limit. The new dispersion relations depend on all three fundamental constants, $G$, $c$ and $\hbar$, and their chosen form gives rise to a new mass-dependence in the phenomenologically significant length scale of the theory. This reduces to the Compton radius for extreme low-mass objects, but corresponds to the Schwarzschild radius in the opposite extreme. We therefore interpret the additional terms in the modified dispersion relations as representing the self-gravity of the quantum wave packet in the semi-Newtonian picture.

In addition, our work suggests a natural definition for the ``Compton temperature" of a fundamental particle. This bears the same relation to the Compton wavelength as does the Hawking temperature of a black hole to its Schwarzschild radius, and may be interpreted as the temperature associated with the particle's rest mass. In the extended de Broglie theory, the two lines, plotted as functions of mass on a log-log plot, define asymptotes to a unified temperature curve, which interpolates between the particle and black hole regions. In this formulation, a Planck mass object is represented by a unique quantum state, whose de Broglie wavelength, Compton wavelength and Schwarzschild radius are equal to the Planck length, and whose temperature is equal to the Planck temperature.  

To summarize: The pedagogical discussions in Secs. \ref{Sec.2}-\ref{Sec.3} aim to clarify the physical assumptions and mathematical foundations on which the extended de Broglie theory is based, while Secs. \ref{Sec.4}-\ref{Sec.6} consider its detailed predictions. The discussion in Sec. \ref{Sec.7} sets the theory proposed in Secs. \ref{Sec.4}-\ref{Sec.6} in the context of the existing literature on modified quantum theories, which attempt to incorporate gravitational effects. Though we make no claims for the quantitative validity of this theory, originally proposed in \cite{Lake:2015pma}, it is hoped that the considerations raised therein, and in the (expanded) discussion presented here, will help stimulate debate on the important open questions in black hole physics: ``Which quantum theory must be reconciled with gravity? (And what does it mean for black holes?)". 






\begin{center}
{\bf Acknowledgments}
\end{center}

ML thanks Bernard Carr and Tiberiu Harko for interesting and helpful discussions.

\renewcommand{\theequation}{A-\arabic{equation}}
\setcounter{equation}{0}  
\section*{Appendix: The postulates of canonical quantum mechanics} \label{Appendix A}


Isham \cite{Ish95} lists four postulates for canonical QM:

\begin{enumerate}

\item The predictions of results of measurements made on an otherwise isolated system are probabilistic in nature. In situations where the maximum amount of information is available, this probabilistic information is represented mathematically by a vector $\ket{\psi}$ in a complex Hilbert space $\mathcal{H}$ that forms the {\it state space} of the theory. In so far as $\ket{\psi}$ gives the most precise predictions that are possible, this vector is to be thought of as the mathematical representation of the physical ``state" of the system.

\item The observables of the system are represented mathematically by self-adjoint operators that act on the Hilbert space $\mathcal{H}$. 

\item If an observable quantity $A$ and a state are represented, respectively, by the self-adjoint operator $\hat{A}$ and the normalized vector $\ket{\psi} \in \mathcal{H}$, the expected result of measuring $A$ is 
\begin{eqnarray} \label{exp_val}
\braket{A}_{\psi} = \braket{\psi|\hat{A}|\psi}.
\end{eqnarray}
(In order to comply with the conventions of standard probability theory, the word ``average" is best reserved for the average of an actual series of measurements. When referring to the ``average" predicted by the mathematical formalism, it is more appropriate to use the phrase ``expected result" or ``expected value".)

\item  In the absence of any external influence (i.e., in a {\it closed} system), the state vector $\ket{\psi}$ changes smoothly in time $t$ according to the time-dependent Schr{\" o}dinger equation
\begin{eqnarray} \label{Schrod_Ish}
i\hbar \frac{\partial \ket{\psi}}{\partial t} = \hat{H}\ket{\psi},
\end{eqnarray}
where $\hat{H}$ is a special operator known as the Hamiltonian.
\end{enumerate}
These four postulates define the general framework within which it has so far been possible to describe \emph{all} canonical QM systems \cite{Ish95}. We note that Postulate 4 is {\it equivalent} to assuming that the operators representing physical observables are defined by analogy with their classical equivalents, together with the assumption that the usual de Broglie relations (\ref{deBroglie}) hold. If these relations are changed, but we retain the analogous classical/quantum formulae, the dispersion relation for the de Broglie waves, and hence the EOM for the quantum system, are also altered. However, this does {\it not} imply any change to the first three postulates, or alter the state space structure of the theory and its associated (probabilistic) interpretation.


\end{document}